\documentstyle[aps,prd,preprint,eqsecnum,epsf]{revtex}
\input epsf.tex

\newcommand{\be}{\begin{equation}}
\newcommand{\ee}{\end{equation}}
\newcommand{\ba}{\begin{eqnarray}}
\newcommand{\ea}{\end{eqnarray}}
\newcommand{\bas}{\begin{eqnarray*}}
\newcommand{\eas}{\end{eqnarray*}}
\newcommand{\hsp}{\hspace{.5cm}}

\newcommand{\tr}{ {\rm{Tr}}}
\newcommand{\R}{ {\cal{R}}}
\newcommand{\la} { {\lambda}}

\tightenlines

\begin{document}

\draft

\title{Anomalies, counterterms 
and the ${\cal N} =0$ Polchinski-Strassler solutions}
\author{Marika Taylor-Robinson
\thanks{email: M.M.Taylor-Robinson@damtp.cam.ac.uk}}
\address{Department of Applied Mathematics and
      Theoretical Physics, \\ Centre for Mathematical Sciences,
      \\ Wilberforce Road, Cambridge CB3 0WA, United Kingdom }
\date{\today}

\maketitle

\begin{abstract}
{The singularity structure of many IIB supergravity solutions
  asymptotic to $AdS_5 \times S^5$ becomes clearer when one considers
  the full ten dimensional solution rather than the dimensionally
  reduced solution of gauged supergravity. It has been shown that all
  divergences in the gravitational action of the dimensionally
  reduced spacetime can be removed by the addition of local 
  counterterms on the boundary. Here we attempt to formulate the
  counterterm action directly in ten dimensions for a particular class of
  solutions, the ${\cal N} = 0$ Polchinski-Strassler solutions, which
  are dual to an ${\cal N} =4$ SYM theory perturbed by mass terms for
  all scalars and spinors. This involves constructing
  the solution perturbatively near the boundary. There is a 
  contribution to the Weyl anomaly from the mass terms (which break 
  the classical conformal invariance of the action). The coefficient 
  of this anomaly is reproduced by a free
  field calculation indicating a non-renormalisation theorem inherited
  from the ${\cal N} =4$ theory. We comment on the structure of 
  the full solutions and their construction from uplifting 
  particular $ {\cal N} = 0$ flows in five dimensions. 
}
\end{abstract}

\pagebreak

\section{Introduction}
\label{sec:intro}

The general form of the 
Maldacena conjecture \cite{Ma}, \cite{GKP}, \cite{W1}, \cite{A}
asserts that there is an equivalence between a gravitational theory in
a $(d+1)$-dimensional anti-de Sitter spacetime and a field
theory in a $d$-dimensional spacetime. 
An interesting consequence of the Maldacena conjecture
is the natural definition of the gravitational action for asymptotically
anti-de Sitter spacetimes without reference to a background \cite{BK},
\cite{KLS}. 

The calculation of the gravitational action has a long history,
particularly in the context of black hole thermodynamics \cite{GH}. One
difficulty that has always plagued this approach is that the gravity
action diverges. The traditional approach to this problem is to use a
background subtraction whereby one compares the action of a spacetime
with that of a reference background, whose asymptotic geometry matches
that of the solution is some well-defined sense. However, this
approach breaks down when there is no appropriate or obvious
background. 

The AdS/CFT correspondence tells us that if, as we expect, the dual
conformal field theory has a finite partition function, then 
we must be able to remove the divergences of the gravitational action
without background subtraction. The framework for achieving this is by
defining local counterterms on the boundary \cite{W1}, \cite{LT},
\cite{HK}, \cite{BK}, \cite{EJM}. For example consider 
the Einstein action in $(d+1)$ dimensions
\be
S = - \frac{1}{16 \pi G_{d+1}} \int_{M} d^{d+1}x \sqrt{g}
( \R + d(d-1) l^2)  - \frac{1}{8 \pi G_{d+1}} \int_{N} \sqrt{\gamma} K
 \label{bulk}
\ee
where $G_{d+1}$ is the Newton constant and $\R$ is the Ricci scalar.
As usual a boundary term must be included for the equations of
motion to be well-defined \cite{GH}, 
with $K$ the trace of the extrinsic curvature of the $d$-dimensional
boundary $N$ embedded into the $(d+1)$-dimensional manifold $M$. Then
provided that the metric near the conformal boundary can be expanded
in the asymptotically anti-de Sitter form
\be
ds^2 = \frac{dr^2}{l^2 r^2} + \frac{1}{r^2} \gamma_{ij} dx^{i} dx^{j},
\label{asym}
\ee
where in the limit $r \rightarrow 0$ the metric $\gamma$ is
non-degenerate, we may remove divergent terms
in the action by the addition of a counterterm action dependent only
on $\gamma$ and its covariant derivatives of the form \cite{BK},
\cite{EJM} 
\be
S_{\rm{ct}} = \frac{1}{8 \pi G_{d+1}} \int_{N} d^dx \sqrt{\gamma}
\left ( (d-1) l + \frac{1}{2 (d-2) l} R(\gamma) + .. \right ) \label{ct} \\
\ee
$R(\gamma)$ and $R_{ij}(\gamma)$ are the Ricci scalar and the Ricci
tensor for the boundary metric respectively. Combined these
counterterms are sufficient to cancel divergences for $d \le 4$, with
a number of exceptions. Firstly, in even dimensions $d = 2n$ one has
logarithmic divergences in the partition function which can be related to
the Weyl anomalies in the dual conformal field theory \cite{HK}.  
Secondly, if the boundary metric becomes degenerate
one can no longer remove divergences by counterterm
regularisation \cite{MMT1}; this is a manifestation of the fact that the dual
conformal field theory does not have a finite partition function in
the degenerate limit. Thirdly, if there are gauge and scalar 
fields on $M$ additional counterterms may be needed to regulate the
action \cite{MMT2}. 

Work on anomalies and counterterms has so far been in the context 
of gauged supergravity theories; that is, we have worked with the
spherical reduction of a ten or eleven dimensional
supergravity solution rather than in the higher dimension. 
Given the interest in extensions of the Maldacena conjecture
to non-conformal theories \cite{W2}, \cite{R1},
\cite{C1}, \cite{R2}, \cite{Cv1}, \cite{Cv2}, \cite{K1}, \cite{M1},
\cite{K2}, \cite{M2}, \cite{KS1}, \cite{G1}, \cite{KS2}, \cite{GPPZ},
\cite{PS}, \cite{Ar1}, \cite{Ar2}, \cite{Ar3}, \cite{Ar4}, \cite{Ar5},
\cite{Ar6}, \cite{Ar7}, particularly to those which
exhibit confinement and condensates, it would be nice
to understand counterterm regularisation of the action for the dual
supergravity backgrounds. For such bulk solutions, one expects to
have counterterms and anomalies related to, for example, the masses
which one has switched on in the gauge theory, in addition to
(\ref{ct}). 

Such solutions are better interpreted in ten (or
eleven) dimensions rather than in the context of some $d$-dimensional
gauged supergravity. This was particularly manifest in the discussions
of \cite{PW}, \cite{Wa1}, \cite{Wa2} 
 where the true IR structure of the GPPZ ${\cal N} =1$
flow \cite{GPPZ} only became manifest when one uplifted to ten
dimensions. Another class of supergravity solutions for which the
ten-dimensional interpretation is essential are the duals of
fractional branes on conifolds \cite{Ar1}, \cite{Ar2}, \cite{Ar3},
\cite{Ar4}, \cite{Ar5}, \cite{Ar6}, \cite{Ar7}. 
So we would like to work with the higher-dimensional solution and 
formulate the corresponding counterterm actions and anomalous
divergences in terms of the higher-dimensional fields. This will shed
light on the nature of the warped compactification and the angular
dependences of the ten-dimensional fields. 

In this paper we will consider the perturbative expansion of one class
of supergravity solutions which are dual to non-conformal gauge
theories, and determine the corresponding counterterms and anomalies. 
Starting from the duality between ${\cal N} = 4$ super
Yang-Mills (SYM) and IIB string theory on $AdS_5 \times S^5$, we 
perturb by the addition of mass terms in the gauge theory which
preserve either less or no supersymmetry. At first sight such a
perturbation appears to produce a spacetime with a naked singularity
in which even basic quantities are incalculable. 

However, Polchinski and Strassler \cite{PS} suggested that the 
naked singularity is replaced by an expanded brane source. This is because 
the mass perturbation in the field theory corresponds to a 3-form 
perturbation in the bulk. The basic idea is that the 
field strength couples to the D3-brane and so the
D3-branes polarise into D5-branes with worldvolume $R^4 \times S^2$ by
the Myers dielectric effect \cite{My}. The background metric for large
distances is dominated by the D3-brane charges but as one moves
towards the IR one starts probing near the D5 brane and so the metric should
be dominated by the D5 branes. The work of \cite{PW}, \cite{Wa1},
\cite{Wa2} showed that the
interpretation is rather more subtle than this: by partially uplifting the GPPZ
flow \cite{GPPZ} to ten dimensions, which gives the equal mass
Polchinski-Strassler solution, they found that if one approaches 
the IR from a generic direction in the sphere one sees a 7-brane. It
is only when one approaches from a direction which is consistent with 
having an IR vacuum in the field theory that one sees five branes. 

Explicit construction of the Polchinski-Strassler supergravity
solutions in ten dimensions is difficult for a number of reasons. 
If one starts with a solution of ${\cal N} =2$ gauged supergravity in
five dimensions and tries to uplift the solution using \cite{NV},
\cite{PW2}, \cite{PW3}, \cite{CLPST} then
the metric on the sphere is so warped that explicit analytic
construction of all ten-dimensional fields is very complex and has not
so far been achieved. However, if one starts from the ten-dimensional
perspective and works with the IIB equations directly, all bosonic fields
are switched on which makes solving even the first order Killing
spinor equations very complex \cite{GrP}. Switching on a mass
perturbation breaks down the R invariance of the field theory and hence
the isometry group of the supergravity solution; all fields acquire an
angular dependence. 

We can simplify the problem a little by trying to preserve the largest
possible isometry group when we make the perturbation. To understand
this, let us consider the ${\cal N} = 4$ theory in the language of
${\cal N} = 1$ supersymmetry; it consists of a vector multiplet $V$
and three chiral multiplets $\Phi_a$ in the adjoint of the gauge
group. We can partially break supersymmetry by adding 
mass terms of the form 
\be
m_{ab} \rm{Tr} (\Phi_a \Phi_b)
\ee
to the superpotential. Such terms break the R-symmetry; this is
manifest since they transform in the $\bf{\bar{6}}$ of the $SU(3)$
flavour symmetry subgroup of the full $SO(6)$. However, if the matrix
is diagonal with three equal eigenvalues, then the 
the resulting ${\cal N} = 1^{\ast}$ theory retains an $SO(3)$ invariance. 
This was exploited in \cite{PW}: writing
the metric in a manifestly $SO(3)$ symmetric form was helpful in
displaying the IR behaviour.

However, even determination of the linearised ten-dimensional 
solution in this more symmetric 
case where we switch on equal mass perturbations 
required a great deal of calculation \cite{MF}. To calculate
the counterterms and anomalies we will need to go beyond the
linearised solution: in effect we need the solution to second order
near the boundary at infinity. We find it therefore calculationally 
easier to make a
further simplification: we consider bulk solutions in which a residual
$[SO(3)]^2$ isometry group is preserved. To preserve this symmetry
there is a price: we have to switch on a mass for the gluino and
break the remaining supersymmetry. 

We do not expect that this breaking of supersymmetry will manifest
itself in the asymptotic solution: after all, the mass perturbation is
relevant and so is a subleading effect in the near boundary solution. 
However, there are a number of
issues we will need to address in the absence of supersymmetry,
particularly whether vacua of the type considered in \cite{PS} exist
even at the classical level and whether they are stable. 

The plan of this paper is as follows. In \S \ref{two} we set up the
framework for constructing ${\cal N} =0$ Polchinski-Strassler
solutions in type IIB supergravity. In \S \ref{three} we
perturbatively solve the equations of motion about the boundary to
adequate order to determine all (UV) divergences in the action. This is
computationally quite intensive and the reader may wish to skip over
the details to \S \ref{four} where we summarise the resulting 
anomalous and divergent terms in the action and discuss how to define
an appropriate counterterm action. 

In \S \ref{five} we consider the action of $SL(2,R)$ on the solutions we
have constructed. This enables us to construct more general solutions
in which the corresponding UV complex coupling and masses in the 
field theory take general values. 

In \S \ref{six} we consider the dual field theory interpretation of
the solutions we have constructed. This involves a discussion of the
appropriate form for the scalar potential in the ${\cal N}=0$ theory
and the corresponding vacua. 
In \S \ref{seven} we discuss five-brane probes moving in the bulk
solution we have constructed. This enables us to fix the normalisation
between the masses in the field theory and the perturbations we have
switched on in the bulk. 

In \S \ref{eight} we show that the anomalous terms in the action
resulting from switching on mass terms are exactly reproduced by a
weak coupling (free field) calculation. This is indicative of a
non-renormalisation theorem inherited from the ${\cal N} =4$ theory to
which it flows in the UV.

In \S \ref{nine} we give some general comments about the full
Polchinski-Strassler solutions. We discuss how one might use the
asymptotic form of the ${\cal N} =0$ solution discussed here to
construct the full solution. The most tractable method is probably to
uplift the corresponding five-dimensional solution which, though not
supersymmetric, should be straightforward to construct. We comment on
the unequal mass solutions and provide further evidence for the mass
anomaly being reproduced by a free field calculation. We also discuss 
the running of the coupling in the dual theory.

\section{Field equations and asymptotic solutions} \label{two}
\noindent

The IIB equations can be derived from the Einstein frame action
\cite{BBO}
\ba
S &=& \frac{1}{2 \kappa^2} \int d^{10}x \sqrt{-G} R - \frac{1}{4\kappa^2}
\int \left ( d\Phi \wedge \ast d\Phi + e^{2\Phi} dC \wedge \ast dC + \right.
\label{ac1} \\
&& \left. g e^{-\Phi} H_3 \wedge \ast H_3 + g e^{\Phi} \bar{F}_3 \wedge \ast
\bar{F}_3 + \frac{1}{2} g^2 \bar{F}_5 \wedge \ast \bar{F}_5 + g^2 C_4
\wedge H_3 \wedge F_3 \right ), \nonumber
\ea
supplemented by the self-duality condition
\be
\ast \bar{F}_5 = \bar{F}_5.
\ee
The other fields are defined as 
\ba
\bar{F}_3 &=& F_3 - C H_3, \hspace{5mm} F_3 = dC_2, \nonumber \\ 
\bar{F}_5 &=& F_5 - C_2 \wedge H_3, \hspace{5mm} F_5 = dC_4.
\ea
Note that $2\kappa^2 = (2 \pi)^7 \alpha'^4 g^2$.
The field equations are \cite{Sch}, \cite{PH}
\ba
\nabla^2 \Phi &=& e^{2 \Phi} \partial_{M} C \partial^{M} C -
\frac{g}{12} e^{-\Phi} H_{MNP} H^{MNP} + \frac{g}{12} e^{\Phi}
\bar{F}_{MNP} \bar{F}^{MNP}, \nonumber \\
\nabla^{M}(e^{2\Phi} \partial_{M}C) &=& -\frac{g}{6} e^{\Phi} H_{MNP}
\bar{F}^{MNP}, \nonumber \\
d \ast (e^{\Phi} \bar{F}_3) &=& g F_{5} \wedge H_{3}, \label{feq} \\
d \ast (e^{-\Phi} H_{3} - C e^{\Phi} \bar{F}_3) &=& - g F_5 \wedge F_3,
\nonumber \\
d \ast \bar{F}_5 &=& - F_3 \wedge H_3, \nonumber
\ea
whilst the Einstein equations are 
\ba
R_{MN} &=& \frac{1}{2} \partial_{M} \Phi \partial_{N} \Phi +
\frac{1}{2} e^{2\Phi} \partial_{M} C \partial_{N}C + \frac{g^2}{96} 
\bar{F}_{MPQRS} \bar{F}_{N}^{\hspace{2mm}PQRS} \nonumber \\
&& + \frac{g}{4} (e^{-\Phi} H_{MPQ} H_{N}^{\hspace{2mm} PQ} + e^{\Phi}
\bar{F}_{MPQ} \bar{F}_{N}^{\hspace{2mm} PQ}) \\
&& - \frac{g}{48} (e^{-\Phi} H_{PQR} H^{PQR} + e^{\Phi}
\bar{F}_{PQR} \bar{F}^{PQR})G_{MN}. \nonumber
\ea
We use indices $\lbrace M,N \rbrace$ in ten dimensions. The Bianchi
identities are 
\ba
d\bar{F}_3 &=& -dC \wedge H_3, \\
d\bar{F}_5 &=& - F_3 \wedge H_3. \nonumber
\ea
As is by now well-known these equations of motion admit a class of
solutions 
\ba
ds^2 &=& H^{-1/2} \eta_{\mu\nu} dx^{\mu} dx^{\nu} + 
H^{1/2} dx^{m} dx^{n}, \nonumber \\
e^{\Phi} &=& g, \hspace{5mm} C = \frac{\theta}{2\pi}, \label{bas} \\
\bar{F}_5 &=& d\chi_4 + \ast d \chi_4, \hspace{5mm} 
\chi_4 = - \frac{1}{gH} dx^{0}
\wedge dx^1 \wedge dx^2 \wedge dx^3, \nonumber 
\ea
with both $g$ and $\theta$ constant and the three-form fields
vanishing. Here the indices $\lbrace \mu, \nu \rbrace$ run between
$0,..,3$ whilst $\lbrace m, n \rbrace = 4,..,9$. $H$ is an arbitrary
harmonic function. For $AdS_5 \times S^5$, 
\be
H = \frac{R^4}{x^4}, \hspace{5mm} R^4 = 4 \pi g N \alpha'^2. 
\ee
We will now look for a solution near the boundary $x \rightarrow
\infty$ in which we switch on three-form fields. For ease of notation
we will set $g = R = 1$ during the calculations and reinstate these
factors at the end. We choose
the following coordinate system for the leading order metric:
\be
ds^2 = \frac{\gamma^0_{\mu\nu}}{r^2} dx^{\mu} dx^{\nu} +
\frac{dr^2}{r^2} + d\theta^2 + \sin^2\theta (d\phi^2 + \sin^2\phi
d\psi_1^2) + \cos^2 \theta (d\chi^2 + \sin^2\chi d\psi_2^2), \label{met}
\ee
where the range of $\theta$ is $\pi/2$ and the other angular
variables take their usual range. Note that $r = 1/x$. $\gamma^0$ is
the background for the dual theory and is taken to be arbitrary.

The reason for the unconventional 
choice for the metric on the $5$-sphere will soon become apparent: we 
find it convenient to use a metric in which an $SO(3)^2$
symmetry group is manifest since our ansatz will preserve this
symmetry group. This $SO(3)^2$ symmetry group is related
to the invariance of the dual ${\cal N} = 0$
theory when all four fermionic mass perturbations are equal. 
 
To obtain the series expansion of the Polchinski-Strassler solution
about the boundary $r=0$ we
need to switch on three-form perturbations . From the approximate solutions
given in \cite{PS} and in \cite{MF}, we know that we should take the leading
order perturbation of the fields to be
\be
\Phi = O(r^2), \hspace{5mm} \bar{F}_3, H_3 = O(r).
\ee
The power of the latter ensures that the solution corresponds to
switching on a non-normalisable relevant perturbation.

We now have to use an appropriate ansatz for the angular
dependence of the three-form fields. In \cite{PS}, the leading order
form form the three-form perturbations was given for the general case
in terms of antisymmetric imaginary-self-dual tensors on $R^6$
satisfying 
\be
\ast_6 T_{mnp} = - i T_{mnp}, \label{ten}
\ee
where $x^m$ are coordinates on $R^6$; it is convenient for what
follows to set $m=4,9$.
Given that we want to write the fluxes explicitly in terms of
coordinates on the sphere it is more convenient to proceed as follows. We are
going to present the explicit details of the calculation for 
the special case for which the RR scalar $C$ vanishes identically and hence
\be
\bar{F}_3 \equiv F_3, \hspace{5mm} F^{MNP} H_{MNP} \equiv 0. \label{con1}
\ee
We will discuss the more general case in a
later section; it is more complicated in details, though
not in principle.  

Motivated by the form of the angular metric in (\ref{met})
there is now a natural choice of ansatz for the leading order
three-form perturbations:
\ba
H_{3} &=& d \lgroup b(x^{\mu},\theta) r \sin\phi d\phi \wedge d\psi_1
\rgroup \equiv d \lgroup b(x^{\mu},\theta) r d\Omega_1 \rgroup, \\
F_{3} &=& d \lgroup c(x^{\mu},\theta) r \sin\chi d\chi \wedge d\psi_2
\rgroup \equiv d \lgroup c(x^{\mu},\theta) r d\Omega_2 \rgroup, \nonumber
\ea
which manifestly satisfies the condition (\ref{con1}). This ansatz
preserves an $[SO(3)]^2$ subgroup of the original $SO(6)$
symmetry group of the sphere and, in addition, evidently satisfies
(\ref{con1}). We will demonstrate explicitly in \S \ref{six} that this
perturbation corresponds to switching on masses for all of the
fermions in the dual theory. 

Switching on such a perturbation will necessarily induce corrections
in the metric and other fields. We are going to look for series
solutions expanded around the boundary; this will involve expanding
the three form perturbations in powers of $br$ and $cr$.  
Given such an ansatz for the three-form fields, the natural choice of
ansatz for the metric expansion is the following:
\be
ds^2 = \frac{\gamma_{\mu\nu}}{r^2} dx^{\mu} dx^{\nu} 
+ \frac{dr^2}{r^2} + G_{i}(r,\theta,x^{\mu}) e^{i} e^{i} + ... ,
\label{m0} 
\ee
where $e^i$ is the sphere vielbein which follows from
(\ref{met}). The ellipses denote ``cross-terms'' in the metric such as
$G_{r\theta}$, which will not be needed
to first order; we will defer discussion of these to later sections.

Following the approach of \cite{HK}, \cite{GL} we will expand $\gamma$
as
\be
\gamma_{\mu \nu} = \gamma^{0}_{\mu \nu} (x^{\rho}) +
\gamma^{2}_{\mu \nu} (x^{\rho}, \theta) r^2 + 
\gamma^{4}_{\mu \nu} (x^{\rho}, \theta) r^4
+ h^{4}_{\mu \nu} (x^{\rho}, \theta) r^4 \ln r + ... \label{m1}
\ee
This ansatz for the metric is the natural generalisation of that given
in \cite{FG}, \cite{GL}. We expect that the series breaks
down at higher order but the terms in (\ref{m1}) should be adequate to
determine all divergences in the action. 

For any solution for which the metric is a warped product asymptotic
to $AdS_5 \times S^5$ the solution should really be interpreted in ten
dimensions to understand the IR behaviour. This means that 
we should work out the series expansion and
counterterms to the action in ten dimensions. The example considered
here is the first of many where one would have to do this. Other
examples include continuous distributions of D3-branes 
\cite{R1}, \cite{R2}, \cite{Cv1}, \cite{Cv2}, \cite{FGPZ};
uplifts of flows such as \cite{GPPZ}, \cite{PW}, \cite{Wa2}, \cite{PW2},
\cite{DGPW}, \cite{PZ}; branes on conifolds \cite{Ar1}, \cite{Ar2},
\cite{Ar3}, \cite{Ar4}, \cite{Ar5}, \cite{Ar6}, \cite{Ar7}. 

Actually we should be a little careful about making an
expansion of the form (\ref{m1}). 
According to \cite{FG}, \cite{GL}, such an expansion always
exists if a $(d+1)$-dimensional Einstein manifold $M$ with negative
cosmological constant has a regular conformal boundary $N$ in the
sense of Penrose \cite{P}. This is manifestly not going to be 
the case here. The full ten-dimensional metric is not Einstein of
negative curvature and furthermore will have a degenerate conformal
boundary. However we could still dimensionally reduce our
ten-dimensional solution to a solution of gauged supergravity in five
dimensions using a warped ansatz to take account of the 
non-trivial angular dependence. 
Hence we expect that a well defined expansion will
exist and that demanding that this expansion is well defined as $r
\rightarrow 0$ will fix the angular dependence of and impose other 
conditions on the matter fields induced on $N$. 

An equivalent way of looking at this is to say that we are going to have to
derive the boundary conditions for a metric to be asymptotically $AdS_5
\times S^5$. This leads to natural extensions of the work of Henneaux and
Teitelboim \cite{HT} on asymptotically AdS spacetimes and 
of Hawking \cite{H} on boundary conditions for gauged supergravity theories. 

We still have to decide how we are going to expand the functions $G_i$
in (\ref{m0}). Given the form of the expansion of the four-dimensional
part of the metric (\ref{m1}) it is reasonable to take the 
expansion of $G^i$ to be
\be
G_{i}(r,\theta,x^{\mu}) = 1 + h^{i}(\theta,x^{\mu}) r^2 +
j^{i}(\theta,x^{\mu}) r^4 + k^{i}(\theta,x^{\mu}) r^4 \ln r +..., \label{m2}
\ee
where by the symmetry of our ansatz $(h^i,j^i,k^i)$ satisfy $h^{\phi} =
h^{\psi_1}$, $h^{\chi} = h^{\psi_2}$ and so on. We expect that the
field equations will explicitly determine the angular dependence of
each of these terms; the $x^{\mu}$ dependence will also be determined
in terms of derivatives of 
the functions $b$ and $c$ and the curvature of the metric $\gamma^0$. 
Finally, we should expand the dilaton (and later the RR scalar also) as 
\be
\Phi = \phi^{(2)}(x^{\rho}, \theta) r^2 + \phi^{(4)}(x^{\rho}, \theta) r^4 +
\varphi^{(4)}(x^{\rho}, \theta) r^4 \ln r + ....,
\ee
where the first order term has vanished since we have taken $g=1$. 

\section{Perturbative solution of field equations} \label{three}
\subsection{First order solution} 
\noindent

Following the approach of \cite{HK} we are going to solve first for
the leading order corrections to the metric and other fields around
the boundary. This means that we should keep the leading order terms 
in the three-form, proportional to the perturbations $b$ and $c$, and
we should keep the corrections of order $r^2$ in all the other
fields. 

Given the ans\"{a}tze for the fields, by matching powers of $r$ we
can work out in which order we need to solve the field equations. It
turns out that we should first solve the coupled equations for the
three-form perturbations. Once we have solved these equations we can
substitute into the dilaton (and RR scalar) equations of motion and
determine their perturbations. Finally, we should solve the coupled Einstein
equations and five-form equations of motion for the $r^2$
perturbations of the metric and five-form. 

So let us first solve the equations for the three-form fields
\ba
d \ast ( e^{\Phi} F_3) &=& F_5 \wedge H_3, \label{3form} \\
d \ast ( e^{-\Phi} H_3) &=& - F_5 \wedge F_3, \nonumber
\ea
which to leading order are 
\ba
3 \frac{\cos^2 \theta}{\sin^2 \theta} b(x^{\mu}, \theta) -
\partial_{\theta} \lgroup \frac{\cos^2 \theta}{\sin^2 \theta}
\partial_{\theta} b(x^{\mu}, \theta) \rgroup 
&=&  - 4 \partial_{\theta} c(x^{\mu}, \theta), \label{cpe} \\
3 \frac{\sin^2 \theta}{\cos^2 \theta} c(x^{\mu}, \theta) -
\partial_{\theta} \lgroup \frac{\sin^2 \theta}{\cos^2 \theta}
\partial_{\theta} c(x^{\mu}, \theta) \rgroup 
&=&  4 \partial_{\theta} b(x^{\mu}, \theta). \nonumber
\ea
The solution to these equations
\ba
b(x^{\mu}, \theta) &=& b \sin^3 \theta, \label{fa} \\
c(x^{\mu}, \theta) &=& b \cos^3 \theta, \nonumber
\ea
where the $x^{\mu}$ dependence of $b$ is implicit and suppressed in
most of what follows, and $b$ is for the
moment arbitrary. We should note here that we can also find a
leading order solution to (\ref{3form}) which is
\ba
H_{3} &=& d (a \cos^3 \theta d \Omega_2); \label{fb} \\
F_{3} &=& - d (a \sin^3 \theta d\Omega_1), \nonumber 
\ea
where again $a$ is an arbitrary function of $x^{\mu}$. However, it is
evident from the RR scalar field equation that we will not be able to
switch on both $a$ and $b$ simultaneously without switching on a
perturbation of $C$. Put another way, $ab$ acts as a source for the RR
scalar. We will find it convenient to discuss first the
case in which $a=0$ and then extend our solutions to the more general
case in which $a$ and the RR scalar are non-zero.
 
In \cite{PS} and \cite{MF}, it was found convenient to combine the R-R
and NS-NS three-forms into a complex three-form $G_3$ by defining
\be
G_{3} = F_{3} - \hat{\tau} H_3, \label{Gf}
\ee
where as usual
\be
\tau = C + i e^{-\Phi},
\ee
and the hat denotes unperturbed fields. This combination is natural
when one considers just the leading order solution and its first order
correction (directly equivalent to the linearised solutions in 
\cite{PS}, \cite{MF}) because of the symmetry between $F_3$ and $H_3$ which is
manifest above. However at higher orders in the radial expansion
there is an asymmetry between the perturbations of 
$F_3$ and $H_3$ caused by the metric and scalar perturbations and we will
find it more convenient to work mostly with these fields rather than $G_3$. 

\bigskip

The dilaton equation of motion depends only on the three-form
perturbations we have just found and on the leading order
metric and five-form:
\be
-4 \Phi(x^{\mu}, \theta) + \frac{1}{\sin^2\theta \cos^2 \theta}
\partial_{\theta} \lgroup \sin^2 \theta \cos^2 \theta
\partial_{\theta} \Phi(x^{\mu}, \theta) \rgroup = 4 b^2 (\sin^2 \theta
- \cos^2 \theta).
\ee
The first correction to $\Phi$ is hence
\be
\Phi(x^{\mu}, \theta) = \frac{1}{4} b^2 (\cos^2 \theta - \sin^2
\theta) r^2.
\ee
We now have all the ingredients needed 
to solve the coupled equations for the corrections to the metric 
and the five-form. Let us write the four-form potential as
\be
C_{4} = C(r,\theta,x^{\mu}) d\eta^{0}_4 + \bar{C}(r,\theta,x^{\mu})
d\Omega_4, \label{5form}
\ee
where we will use the natural shorthand notation
\ba
d\eta^{0}_4 &=& \sqrt{\gamma^0} dx^{0} \wedge dx^1 \wedge dx^2 \wedge
dx^3, \nonumber \\
d\Omega_4 &=& d\Omega_1 \wedge d\Omega_2.
\ea
Then to get the first corrections to the five form 
we should expand the functions in (\ref{5form}) as 
\ba
C &=& - \frac{1}{r^4} - \frac{\alpha^{(2)}}{2r^2}, \\
\bar{C} &=& 4 \int \sin^2 \theta \cos^2 \theta d\theta + \beta^{(2)}
r^2. \nonumber
\ea
It is convenient here to make a further assumption, namely that 
\be
\beta^{(2)} = \frac{1}{2} b^2 \sin^3 \theta \cos^3 \theta, \label{bet1}
\ee
which implies that $\bar{F}_{r\Omega_4} = \bar{F}_{\mu \Omega_4} = 0$ 
and
\be
\bar{F}_{\theta \Omega_4} = \sin^2\theta \cos^2 \theta \lgroup 4 -
\frac{3}{2} b^2 r^2 \rgroup.
\ee
Using one of the constraints arising from the self-duality of the
five-form we find that the vanishing of $\bar{F}_{r\Omega_4}$ implies
the vanishing of $\bar{F}_{\theta \eta_4}$ and hence that
$\alpha^{(2)}$ does not depend on $\theta$.  
Although at this stage there is no justification for this assumption
if we keep $\beta^{(2)}$ arbitrary during the calculation we find that
the $[SO(3)]^2$ symmetry will later force it to take the value (\ref{bet1}). 
Fixing $\beta$ is effectively a coordinate choice 
not a gauge choice; we have already fixed the gauge
in the choice of the ansatz (\ref{5form}).
Reasons for this choice of $\beta$ will be discussed at the end of
this section. Note the simplicity of the ansatz for the five-form for
this symmetric case compared to that given for the linearised solution
in \cite{MF} which corresponds to massless gluinos. 

Given the ansatz for the five-form and the metric, as well as the
solutions we already have for the three-form perturbations, we can
now work out the explicit expansions of the Einstein equations. 
The angular Einstein equations are:
\ba
R_{\theta \theta} &=& 2 h^{\theta} - \frac{1}{2} \tr( (\gamma^0)^{-1}
\gamma^2)_{,\theta \theta} - \frac{1}{2} \sum_{\i \neq \theta}
h^{i}_{,\theta \theta} - \lgroup 2 h^{\phi} - h^{\theta} \rgroup
_{,\theta} \frac{\cos \theta}{\sin \theta} + 
\lgroup 2 h^{\chi} - h^{\theta} \rgroup
,_{\theta} \frac{\sin \theta}{\cos \theta}; \nonumber \\
&=& \frac{b^2}{4} - 4 \sum_{i \neq \theta} h^{i}. \nonumber \\
\frac{R_{\phi \phi}}{\sin^2 \theta} &=& - \frac{\cos\theta}{2 \sin\theta} 
\tr( (\gamma^0)^{-1} \gamma^2)_{,\theta} + h^{\phi} (6 -
\frac{1}{\sin^2\theta}) + h^{\theta} (\frac{1}{\sin^2\theta} -4) -
\frac{1}{2} h^{\phi}_{,\theta\theta} \nonumber \\
&& + h^{\phi}_{,\theta} \lgroup 
\frac{\sin\theta}{\cos\theta} - \frac{2 \cos \theta}{\sin \theta}
\rgroup + \lgroup \frac{1}{2} h^{\theta}_{,\theta} -
h^{\chi}_{,\theta} \rgroup \frac{\cos\theta}{\sin\theta}; \nonumber \\
&=& b^2 (4 \cos^2 \theta - \frac{15}{4}) - 4 \sum_{i \neq \phi} h^{i}.
\nonumber \\
\frac{R_{\chi \chi}}{\cos^2 \theta} &=&  \frac{\sin\theta}{2 \cos\theta} 
\tr( (\gamma^0)^{-1} \gamma^2)_{,\theta} + h^{\chi} (6 -
\frac{1}{\cos^2\theta}) + h^{\theta} (\frac{1}{\cos^2\theta} -4) -
\frac{1}{2} h^{\phi}_{,\theta\theta} \nonumber \\
&& + h^{\chi}_{,\theta} \lgroup 
\frac{2\sin\theta}{\cos\theta} - \frac{\cos \theta}{\sin \theta}
\rgroup - \lgroup \frac{1}{2} h^{\theta}_{,\theta} -
h^{\chi}_{,\theta} \rgroup \frac{\sin\theta}{\cos\theta}; \nonumber \\
&=& b^2 (4 \sin^2 \theta - \frac{15}{4}) - 4 \sum_{i \neq \chi} h^{i}.
\nonumber \\
R_{r \theta} &=& 0; \label{e1} \\
&=& - \tr( (\gamma^0)^{-1} \gamma^2)_{,\theta} - \sum_{\i \neq \theta}
h^{i}_{,\theta} + 2( h^{\theta} - h^{\phi})
\frac{\cos\theta}{\sin\theta} - 2(h^{\theta} - h^{\chi})
\frac{\sin\theta}{\cos\theta}. \nonumber \\
R_{\mu \theta} &=& 0; \nonumber \\
&=& - \frac{1}{2} \sum_{i \neq \theta} h^{i}_{,\mu \theta} +
\frac{1}{2} (\gamma^0)^{\nu \rho} \lgroup \nabla^{0}_{\mu} 
\gamma^2_{\nu \rho} - \nabla^{0}_{\rho} \gamma^2_{\mu\nu} \rgroup_{,\theta} +
\lgroup h^{\theta}_{,\mu} - h^{\phi}_{,\mu} \rgroup
\frac{\cos\theta}{\sin\theta} - \lgroup h^{\theta}_{,\mu} 
- h^{\chi}_{,\mu} \rgroup \frac{\sin\theta}{\cos\theta}. \nonumber 
\ea
In this list, we give first the curvature calculated using the metric
ansatz and second the trace adjusted stress energy tensor calculated
using the fields already found and the ans\"{a}tze for the other fields. 
Here $\nabla^0$ is the covariant derivative associated with
$\gamma^0$. The $SO(3)^2$ symmetry of the
ansatz implies that the $R_{\psi_i\psi_i}$ equations are equivalent to
the $R_{\phi\phi}$ and $R_{\chi\chi}$ equations. Furthermore, the
symmetry of the ansatz for the three-form and five-form is going to
force a symmetry between the $h^{\phi}$ and $h^{\chi}$ perturbations. 

The other Einstein equations are:
\ba
R_{rr} &=& - 2 \sum_{i} h^{i} = 4 \tr ( (\gamma^0)^{-1} \gamma^2) - 2
\alpha^{(2)} - \frac{3}{4} b^2. \nonumber \\
R_{\mu r} &=& \frac{1}{2} b b_{,\mu} = - 
(\gamma^0)^{\nu \rho} \lgroup \nabla^{0}_{\mu} \gamma^2_{\nu \rho} -
\nabla^{0}_{\rho} \gamma^2_{\mu \nu} \rgroup 
- \frac{3}{2} \sum_{i} h^{i}_{,\mu}. \label{e2}
\\
R_{\mu\nu} &=& R^{0}_{\mu\nu} - 2 (\gamma^2)_{\mu\nu} + 
\tr ( (\gamma^0)^{-1} \gamma^2) (\gamma^0)_{\mu\nu} - \frac{1}{2}
(\gamma^2_{\mu\nu})_{,\theta}(\frac{\cos\theta}{\sin\theta} - \frac{\sin\theta}{\cos\theta}) \nonumber \\
&& - (\gamma^2_{\mu\nu})_{,\theta \theta} + \sum_{i} h^{i}
(\gamma^0)_{\mu\nu}; \nonumber \\
&=& 4 \tr ( (\gamma^0)^{-1} \gamma^2) (\gamma^0)_{\mu\nu} 
- 4 (\gamma^2)_{\mu\nu} - 2 \alpha^{(2)} (\gamma^0)_{\mu\nu} - \frac{5}{4}
b^2 (\gamma^0)_{\mu\nu}, \nonumber
\ea
where again we give the explicitly calculated Ricci tensor followed by
the stress energy tensor. 
We have already mentioned one constraint arising from the self-duality
of the five-form; the second constraint relates $\bar{F}_{r\eta_4}$ to
$\bar{F}_{\theta \Omega_4}$ and gives the final equation of motion:
\be
\alpha^{(2)} = 2 \tr ( (\gamma^0)^{-1} \gamma^2) - 2 \sum_{i} h^{i} -
\frac{3}{2} b^2. \label{e3}
\ee
The three groups of equations (\ref{e1}), (\ref{e2}) and (\ref{e3})
involve unknowns which depend both on $\theta$ and on $x^{\mu}$. The 
angular dependence of each unknown must be fixed by the consistency
of the ansatz, whilst the $x^{\mu}$ dependence is arbitrary. 
So we can solve the coupled
equations by expanding each variable in spherical harmonics and then
fixing the coefficients. The resulting solution for the metric and
five-form perturbations is
\ba
\gamma^{(2)}_{\mu\nu} &=& -\frac{1}{2} R^{0}_{\mu\nu} + \frac{1}{12} R^{0}
\gamma^0_{\mu\nu} + \frac{5b^2}{48} (\gamma^0)_{\mu\nu}; \nonumber \\
\tr( (\gamma^{0})^{-1} \gamma^2) &=& -\frac{1}{6} R^{0} +
\frac{5b^2}{12}; \nonumber \\
h^{\theta} &=& \frac{b^2}{8}; \\
h^{\phi} = h^{\psi_1} &=& \frac{b^2}{8} (1 - 4 \sin^2 \theta); \nonumber \\
h^{\chi} = h^{\psi_2} &=& \frac{b^2}{8} (1 - 4 \cos^2 \theta); \nonumber \\
\alpha^{(2)} &=& \frac{b^2}{12} - \frac{R^0}{3}.
\ea
The final constraint induced by the field equations 
is that the torsion associated with the metric $\gamma^0$
vanishes, the same constraint as was found in \cite{HK}. 

Let us mention here that we could also make a perturbation such that 
\ba
h^{\theta} &=& \frac{1}{6} \beta (\cos^2 \theta - \sin^2 \theta);
\nonumber \\
h^{\phi} &=& \frac{1}{6} \beta \cos^2 \theta; \hspace{5mm} h^{\chi} =
- \frac{1}{6} \beta \sin^2 \theta; \nonumber \\
G_{\mu\theta} &=& \frac{1}{12} \beta_{,\mu} r^2 \cos\theta \sin\theta;
\hspace{5mm} G_{r\theta} = \frac{1}{6} \beta r \cos\theta \sin\theta;
\\
\bar{F}_{\theta \Omega_4} &=& \sin^2 \theta \cos^2 \theta \left (4 + \beta
r^2 (\cos^2 \theta - \sin^2 \theta) \right); \nonumber \\
\bar{F}_{r \Omega_{4}} &=& \frac{2}{3} \beta r \cos^3\theta \sin^3
\theta; \hspace{5mm} \bar{F}_{r\eta} = \frac{4}{r^5}, \nonumber 
\ea
with the three-form fields vanishing. This is precisely the effect of
rescaling 
\be
\theta \rightarrow \theta + \frac{1}{12} \beta \cos\theta \sin \theta
r^2
\ee
leaving all other coordinates fixed, and hence we can always eliminate
$G_{r\theta}$, $G_{\mu\theta}$ terms to first order with a suitable
choice of $\beta$. We expect, however, that we will need to include
such cross terms in the metric ansatz at higher order in order to find
a solution. Coordinate transformations cannot remove cross terms to
all orders. 

A final comment is that the solution we have found here is closely
related to the (extremal limit of the) linearised solution
presented in \cite{MF}. The difference is that our solution
corresponds to switching on four equal fermion masses rather than
three. We will expand on this in a later section when we discuss the
form for the more general Polchinski-Strassler solutions. 

\subsection{Logarithmic terms}
\noindent

To determine all divergent terms in the action we need to carry on
expanding the fields about the boundary.
It turns out that we need to solve next for the logarithmic corrections 
to the fields, that is, metric corrections of order $r^4 \ln r$ and
corresponding terms in the other fields, rather than for second 
order perturbations to the
fields, meaning metric corrections of order $r^4$. The reason for
this is that radial derivatives of the logarithmic corrections will
contribute to the $r^4$ terms.  

The ansatz for the logarithmic terms in the metric is
contained in (\ref{m1}) and (\ref{m2}). We also need to allow for
cross-terms in the metric; although the field redefinition discussed
above can remove the leading order cross-terms we should 
expect that the following terms in the metric could be present:
\ba
G_{r\theta} &=& r^3 k^{r\theta}(\theta,x^{\mu}) \ln r; \nonumber \\
G_{r \mu} &=& r^3 k^{r}_{\mu} (\theta, x^{\mu}) \ln r; \\
G_{\mu \theta} &=& r^4 k^{\theta}_{\mu} (\theta, x^{\mu}) \ln r.
\nonumber
\ea
These are the only terms allowed by the symmetry of our ansatz. Notice
that we fix the power of $r$ in these terms by demanding 
that the contributions to the Ricci tensor are of the appropriate order. For
simplicity we will anticipate our results and set $k^{r\theta} =
k^{\theta}_{\mu} = 0$ from the start. 

As in the previous section, we need to solve the equations of motion
in the order three-form equations, scalar field equations and
Einstein/five-form equations. So let us solve first the three-form
field equations. Taking the ansatz to be 
\ba
H_{3} &=& d \lgroup b r \sin^3 \theta + \la(\theta,x^{\mu}) r^3 \ln r
\rgroup d \Omega_1; \\
F_{3} &=& d \lgroup b r \cos^3 \theta + \bar{\la}(\theta,x^{\mu}) 
r^3 \ln r \rgroup d \Omega_2, \nonumber
\ea
we find a similar pair of equations to (\ref{cpe}) which admit the
solutions 
\ba
\la(\theta,x^{\mu}) &=& \la \sin^3 \theta; \\
\bar{\la}(\theta,x^{\mu}) &=& \la \cos^3 \theta. \nonumber
\ea
Given these solutions we can solve the dilaton equation
to get the following logarithmic contribution 
\be
\Phi = .... + \frac{1}{2} \la b (\cos^2 \theta - \sin^2 \theta) r^4 \ln
r.
\ee
From the solutions for the three-form we can substitute these into
the coupled five-form and Einstein equations to determine metric and
five-form perturbations in terms of $\lambda$ and $b$. 
The ansatz we will use for the five-form is the following:
\ba
\bar{F}_{\theta \Omega_4} &=& 4 \sin^2 \theta \cos^2 \theta ( 1 + ... -
\frac{3}{4} \la b r^4 \ln r); \\
\bar{F}_{r \eta} &=& (\frac{4}{r^5} + ... + \alpha^{(4)}_{\rm{log}}
\frac{\ln r}{r} ), \nonumber 
\ea
where the ellipses denote the first order perturbations found in the previous
section which are not relevant
here and all other logarithmic terms vanish. Again we are using hindsight to
write down this ansatz; there is no reason a priori 
why all the other components should vanish. Note that the absence of a
$(\ln (r))^2$ term in $\bar{F}_{\theta \eta}$ will require that
$\alpha^{(4)}_{\rm{log}}$ is independent of $\theta$, a condition which
we will find is indeed satisfied by our solution. 

Using this ansatz and the solutions for the three-form perturbations,
we can write the angular Einstein equations as
\ba
R_{\theta \theta} &=& -\frac{1}{2} \tr( (\gamma^0)^{-1} h^4)_{,\theta
  \theta} - \frac{1}{2} \sum_{i \neq \theta} k^{i}_{,\theta \theta} 
 + (k^{\theta} - 2k^{\phi})_{,\theta} \frac{\cos \theta}{\sin \theta} 
 - (k^{\theta} - 2 k^{\chi})_{,\theta} \frac{\sin \theta}{\cos
   \theta}; \nonumber \\
&=& - 4 \sum_{i \neq \theta} k^{i}. \nonumber \\
\frac{R_{\phi \phi}}{\sin^2 \theta} &=&
-\frac{\cos\theta}{2\sin\theta}  \tr( (\gamma^0)^{-1} h^4)_{,\theta}
+ (k^{\phi} - k^{\theta})( 4 - \frac{1}{\sin^2\theta}) - \frac{1}{2}
k^{\phi}_{,\theta \theta} \nonumber \\
&& + (k^{\theta} - 4 k^{\phi} - 2
k^{\chi})_{,\theta} \frac{\cos\theta}{\sin\theta} + k^{\phi}_{,\theta}
\frac{\sin \theta}{\cos \theta}; \nonumber \\
&=& -6 \la b \sin^2 \theta - 4 \sum_{i \neq \phi} k^{i}. \\
\frac{R_{\chi \chi}}{\cos^2 \theta} &=&
\frac{\sin\theta}{2\cos\theta}  \tr( (\gamma^0)^{-1} h^4)_{,\theta}
+ (k^{\chi} - k^{\theta})( 4 - \frac{1}{\cos^2\theta}) - \frac{1}{2}
k^{\chi}_{,\theta \theta} \nonumber \\ 
&& - (k^{\theta} - 4 k^{\chi} - 2
k^{\phi})_{,\theta} \frac{\sin\theta}{\cos\theta} - k^{\chi}_{,\theta}
\frac{\cos \theta}{\sin \theta}; \nonumber \\
&=& -6 \la b \cos^2 \theta - 4 \sum_{i \neq \chi} k^{i}. \nonumber \\
R_{r \theta} &=&  - 2 \sum_{i \neq \theta} k^{i}_{,\theta} + 4
(k^{\theta} - k^{\phi}) \frac{\cos \theta}{\sin\theta} - 4 (k^{\theta}
- k^{\chi}) \frac{\sin \theta}{\cos \theta} - \frac{3}{2} 
 \tr( (\gamma^0)^{-1} h^4)_{,\theta}; \nonumber \\
 &=& 0. \nonumber \\
R_{\mu \theta} &=& (\gamma^0)^{\nu \rho} (\nabla^{0}_{\rho}
h^{4}_{\mu \nu} -  \nabla^{0}_{\mu} h^{4}_{\nu \rho})_{,\theta} 
- \frac{5}{4} \sum_{i \neq \theta} k^{i}_{\mu \theta}; \nonumber \\
&=& 0, \nonumber 
\ea
where as before we have written first the curvature terms and secondly
the trace adjusted stress energy tensor.
The other Einstein equations are 
\ba
R_{rr} &=& -8 \sum_{i} k^{i} - 4 \tr( (\gamma^0)^{-1} h^4); \nonumber
\\
&=& -2 \alpha^{(4)}_{\rm{log}} + 4 \tr( (\gamma^0)^{-1} h^4). \nonumber
\\
R_{\mu r} &=& 2 (\gamma^0)^{\nu \rho} (\nabla^{0}_{\rho}
h^{4}_{\mu \nu} -  \nabla^{0}_{\mu} h^{4}_{\nu \rho}) 
- \frac{5}{2} \sum_{i \neq \theta} k^{i}_{\mu} - 4k^{r}_{\mu}; \nonumber \\
&=& \frac{1}{2} (b \la_{,\mu} + 3 \la b_{,\mu}). \\
R_{\mu \nu} &=& -4 h^{4}_{\mu\nu} - \frac{1}{2} h^{4}_{\mu\nu , \theta
  \theta} - h^{4}_{\mu\nu , \theta} (\frac{\cos \theta}{\sin \theta} -
\frac{\sin \theta}{\cos \theta}) + 2 \sum_{i} k^{i}
\gamma^0_{\mu\nu} + 2 \tr( (\gamma^0)^{-1} h^4) \gamma^0_{\mu\nu};
\nonumber \\
&=& \lgroup -3 \la b - 2 \alpha^{(4)}_{\rm{log}} + 2 \tr(
(\gamma^0)^{-1} h^4) \rgroup \gamma^0_{\mu\nu}. \nonumber
\ea
Here $R^{0}$ is the curvature associated with the metric $\gamma^0$.
The final equation is given by the self-duality condition on the
five-form
\be
\alpha^{(4)}_{\rm{log}} = -2 \sum_{i} k^{i} + 2 \tr(
(\gamma^0)^{-1} h^4) - 3 \la b.
\ee
These equations may again be solved by assuming that the unknowns depend
only on the first harmonics in $\theta$ which reduces most of the
differential equations to algebraic equations. Solving everything but
the $R_{\mu r}$ equation we find the following solutions for the metric
and five-form perturbations
\ba
\alpha^{(4)}_{\rm{log}} &=& 0; \nonumber \\
\tr( (\gamma^0)^{-1} h^4) &=& \frac{3}{4} \la b; \nonumber \\
k^{\theta} &=& \frac{1}{4} \la b; \label{slog} \\
k^{\phi} = k^{\psi_1} &=& \frac{1}{4} \la b (1 - 4 \sin^2 \theta); \nonumber \\
k^{\chi} = k^{\psi_2} &=& \frac{1}{4} \la b (1 - 4 \cos^2 \theta); \nonumber
\ea
The $R_{\mu r}$ equation then determines the logarithmic contribution to
the $G_{\mu r}$ component to be
\be
k^{r}_{\mu} = \frac{1}{16} b \la_{,\mu} - \frac{3}{16} \la b_{,\mu}.
\ee
Only the trace of $h^4$ is fixed by these equations; the rest of $h^4$
will be determined by the equations of motion for the 
non-logarithmic terms at the same radial
order. Note that the trace of $h^4$ does not vanish.
At first sight, this seems a little worrying since it
implies we might end up getting divergences of the form $(\ln \epsilon)^2$ in
the action, where $\epsilon$ is the cut-off length scale; such
divergences do not have a natural interpretation. However it will turn
out that there are other contributions to these divergences and the
net contribution cancels in a non-trivial way. 

\subsection{Second order terms} \label{foura}
\noindent

Given the solutions for the first order and logarithmic perturbations
we can now solve for the $r^4$ corrections to the metric and
corresponding terms.
Once again we should start with the coupled equations for the
three forms. Taking the ansatz to be
\ba
H_{3} &=& d \lgroup b r \sin^3 \theta + \la \sin^3 \theta r^3 \ln r + 
  B(\theta,x^{\mu}) r^3 \rgroup d\Omega_1; \\
F_{3} &=& d \lgroup b r \cos^3 \theta + \la \cos^3 \theta r^3 \ln r + 
  C(\theta,x^{\mu}) r^3 \rgroup d\Omega_2, \nonumber
\ea
and substituting into the field equations, using the explicit values
of the first order terms, we find the following coupled equations:
\ba
-\partial_{\theta} \lgroup \frac{\cos^2 \theta}{\sin^2 \theta}
B_{,\theta} \rgroup + 3 B \frac{\cos^2 \theta}{\sin^2 \theta}
&=& -4 C_{,\theta} + 24 b^3 \sin^4\theta \cos\theta + K \sin^2 \theta
\cos \theta; \label{eq3}
\\
-\partial_{\theta} \lgroup \frac{\sin^2 \theta}{\cos^2 \theta}
C_{,\theta} \rgroup + 3 C \frac{\sin^2 \theta}{\cos^2 \theta}
&=& -4 B_{,\theta} + 24 b^3 \cos^4\theta \sin\theta + K \cos^2 \theta 
\sin \theta, \nonumber
\ea
where 
\be
K = (\Box^{0} -\frac{1}{6} R^{0}) b + 2 \la - \frac{53}{6} b^3.
\ee
Note that there is a contribution here from the radial derivatives of
the logarithmic terms in $F_3$ and $H_3$. This is why we had to work
out the logarithmic corrections to the fields before solving for the
$r^4$ corrections. 
Solving (\ref{eq3}) turns out to be quite subtle. Let us first
substitute in 
\be
B(\theta, x^{\mu}) = \hat{B}(\theta,x^{\mu}) - \frac{1}{2} b^3 \sin^5 
\theta, \hsp C(\theta, x^{\mu}) =
\hat{C}(\theta, x^{\mu}) - \frac{1}{2} b^3 \cos^5 \theta,
\ee
so that 
\ba
-\partial_{\theta} \lgroup \frac{\cos^2 \theta}{\sin^2 \theta}
\hat{B}_{,\theta} \rgroup + 3 \hat{B} \frac{\cos^2 \theta}{\sin^2 \theta}
&=& -4 \hat{C}_{,\theta} + \hat{K} \sin^2 \theta \cos \theta; \label{eqn4}
\\
-\partial_{\theta} \lgroup \frac{\sin^2 \theta}{\cos^2 \theta}
\hat{C}_{,\theta} \rgroup + 3 \hat{C} \frac{\sin^2 \theta}{\cos^2 \theta}
&=& -4 \hat{B}_{,\theta} + \hat{K} \cos^2 \theta \sin \theta, \nonumber
\ea
where now
\be
\hat{K} = (\Box^{0} -\frac{1}{6} R^{0}) b + 2 \la + \frac{1}{6} b^3.
\ee
We now want to argue that a sensible (non-divergent) solution to (\ref{eqn4}) 
exists only when $\hat{K} = 0$. When $\hat{K} \neq 0$ there is
effectively an inhomogeneous perturbation which coincides with the
solution to the homogeneous equations. This will mean that the
resulting solution will be singular at some value of $\theta$ and
since the range of $\theta$ is finite this singularity will result in a
real physical singularity of the metric and other fields. This effect
is directly analogous to resonance when an oscillatory system is
forced at its natural frequency. 

Unfortunately, explicit solutions to (\ref{eqn4}) when $\hat{K} \neq 0$
in terms of elementary functions do not seem to be accessible. The
best way to demonstrate that
the solutions are divergent is to expand $\hat{B}$ and $\hat{C}$ in
harmonics on the $S^5$: the resulting series show  
divergences at $\theta = 0$ or $\theta = \pi/2$. 

\bigskip

When $\hat{K} = 0$, solution of (\ref{eqn4}) proceeds as in 
the previous sections so that the three-form fields are
\ba
H_{3} &=& d \lgroup b r \sin^3 \theta + \la \sin^3 \theta r^3 \ln r + 
  (B - \frac{1}{2}b^3 \sin^2\theta) r^3 \sin^3\theta \rgroup d\Omega_1; \\
F_{3} &=& d \lgroup b r \cos^3 \theta + \la \cos^3 \theta r^3 \ln r + 
  (B - \frac{1}{2} b^3 \cos^2 \theta) r^3 \cos^3 \theta 
\rgroup d\Omega_2. \nonumber 
\ea
$B(x^{\mu})$ is a new field, deriving from the homogeneous part of the
solution to (\ref{eqn4}). It gives rise to a normalisable perturbation
and corresponds to switching on a vacuum expectation value for fermion
bilinears.
We can now substitute directly into the dilaton equation of motion to
get the following solution
\be
\Phi = \left (  \frac{1}{4} b^2 r^2 + (\frac{1}{2} Bb - \frac{1}{144}
R^{0}b + \frac{71}{576} b^4)r^4  + \frac{1}{2} \la b r^4 \ln r \right )
(\cos^2 \theta - \sin^2 \theta). 
\ee
We need to give an ansatz for the metric and for the five-form. As
well as the terms given in (\ref{m1}) and (\ref{m2}), we should also
include the following cross-terms 
\ba
G_{r \theta} &=& j^{r \theta} r^3; \nonumber \\ 
G_{r \mu} &=& j^{r}_{\mu} r^3; \\
G_{\mu \theta} &=& j^{\theta}_{\mu} r^3. \nonumber
\ea
We have derived the powers of $r$ by demanding that the contribution
to the Einstein tensor is of appropriate order. Thus as anticipated we
are being forced to derive the falloff of cross terms in the metric,
or in another words the conditions for a spacetime to be
asymptotically $AdS_5 \times S^5$. 

The appropriate ansatz for the five-form field is then
\ba
\bar{F}_{\theta \Omega_4} &=& 4 \sin^2 \theta \cos^2 \theta \lbrace 1 -
\frac{3b^2 r^2}{8} - \frac{3Bb r^4}{4} + \frac{3 f_{-} r^4}{16}
(\cos^2 \theta - \sin^2 \theta) \nonumber \\
&& + f_{+} r^4 (\frac{3}{16} -
\sin^2\theta \cos^2 \theta) + b^4 r^4 (\frac{9}{32} - \frac{1}{4}
\sin^2\theta \cos^2 \theta) \rbrace. \nonumber \\
\bar{F}_{r \Omega_{4}} &=& \lbrace f_{-} - \la b + f_{+} (\cos^2 \theta
- \sin^2 \theta) \rbrace r^3 \cos^3 \theta \sin^3 \theta. \nonumber \\
\bar{F}_{\mu \Omega_{4}} &=& \frac{1}{4} \lbrace f_{-} + f_{+} (\cos^2 \theta
- \sin^2 \theta) \rbrace_{,\mu} r^3 \cos^3 \theta \sin^3
\theta. \label{eqn5} \\
\bar{F}_{\theta \eta_{4}} &=& \lbrace - f_{-} + \la b - f_{+} (\cos^2 \theta
- \sin^2 \theta) + 4 j^{r \theta} \rbrace \cos\theta
\sin\theta. \nonumber \\
\bar{F}_{r \eta_{4}} &=& \frac{4}{r^5} + (\frac{b^2}{12 r^3} -
\frac{R^0}{3 r^3}) + \frac{\alpha^{(4)}}{r}. \nonumber
\ea
There are various comments to make about this ansatz. Firstly, we need
to introduce two new parameters $\{ f_{-}, f_{+} \}$ to obtain a
general enough form; when we solve the equations of motion these
parameters will however be fixed in terms of $\{ b,B \}$. 

Secondly, in this ansatz we have already partially implemented self-duality;
that is, we have exploited the duality between $\bar{F}_{r
\Omega_{4}}$ and $\bar{F}_{\theta \eta_{4}}$. Note that the term
$\bar{F}_{\mu \Omega_{4}}$ must correspond to a dual term 
$\bar{F}_{r \theta \nu
\rho \sigma}$ which we have not included in our ansatz. However,
keeping track of powers of $r$ we can see that the latter
is higher order, does not contribute to the field equations and can be
ignored in all that follows.  

Thirdly, we have already implemented the $\bar{F}_{5}$ field equation
in this ansatz, namely that 
\be
d (\ast \bar{F}_5) = - F_{3} \wedge H_3.
\ee
This condition determines the relationship between
$\bar{F}_{r\Omega_4}$, $\bar{F}_{\theta \Omega_4}$ and $\bar{F}_{\mu
  \Omega_4}$, using the already determined expressions for $F_3$ and
$H_3$. 

The condition that $F_{5}$ is exact imposes the further constraint 
that 
\be
\partial_{\theta} \alpha^{(4)} = 0,
\ee
since we have found in the previous section 
that there is no logarithmic term in $\bar{F}_{\theta \eta_4}$. 
The final constraint from self-duality requires that
\ba
\alpha^{(4)} &=& 2 \tr( (\gamma^0)^{-1} \gamma^4) - 2 \sum_{i} j^{i} - 3 Bb
+ \frac{3}{4} f_{-} (\cos^2\theta - \sin^2 \theta) \nonumber \\
&& + f_{+} (\frac{3}{4} - 4 \sin^2 \theta \cos^2 \theta) + b^4
(\frac{1301}{1152} - 2 \sin^2\theta \cos^2 \theta) \\
&& + \frac{1}{36} b^2 R^{0} - \frac{1}{4} (R^{0})^{mn} R^{0}_{mn} +
\frac{5}{72} (R^{0})^2. \nonumber  
\ea
Using the fields already determined as well as the ans\"{a}tze for 
the metric and five-form perturbations, we can work out 
the angular Einstein equations which are
\ba
R_{\theta \theta} &=& -\frac{1}{8} \lgroup b (\Box^{0} - \frac{1}{6}
R^{0}) b + (\partial b)^2 \rgroup - \frac{1}{2} 
\tr( (\gamma^0)^{-1} \gamma^4)_{,\theta \theta} - \frac{1}{2} \sum_{i
  \neq \theta} j^{i}_{,\theta \theta} - \frac{1}{2} \la b \nonumber \\
&& + (j^{\theta} - 2 j^{\phi})_{,\theta} \frac{\cos\theta}{\sin\theta} 
- (j^{\theta} - 2 j^{\chi})_{,\theta} \frac{\sin\theta}{\cos\theta}
+ b^4 (- \frac{139}{192} \cos^4\theta + \frac{341}{96} \cos^2 \theta
\sin^2 \theta - \frac{139}{192} \sin^4 \theta);
\nonumber \\
&=& -4 \sum_{i \neq \theta} j^{i} + \frac{3 f_{+}}{2} - 8 f_{+} \sin^2
\theta \cos^2 \theta + b^4 
( \frac{63}{64} \cos^4\theta - \frac{97}{32} \cos^2\theta 
\sin^2\theta + \frac{63}{64} \sin^4 \theta) \nonumber \\
&& - \frac{1}{4} \la b - \frac{1}{8} (\partial b)^2. \nonumber \\
\frac{R_{\phi \phi}}{\sin^2 \theta} &=& - \frac{1}{8} (1 - 4 \sin^2
\theta) \lgroup b (\Box^{0} - \frac{1}{6} R^{0}) b + (\partial b)^2
\rgroup - \frac{\cos \theta}{2 \sin \theta} 
\tr( (\gamma^0)^{-1} \gamma^4)_{,\theta} \nonumber \\
&& + (j^{\phi} - j^{\theta})(4 - \frac{1}{\sin^2\theta})  
- \frac{1}{2} j^{\phi}_{,\theta \theta} 
 + (\frac{1}{2} j^{\theta} - 2 j^{\phi} - j^{\chi})_{,\theta} 
\frac{\cos\theta}{\sin\theta} 
+  j^{\phi}_{,\theta} \frac{\sin\theta}{\cos\theta} \nonumber \\
&& + b^4 (- \frac{139}{192} \cos^4\theta + \frac{91}{96} \cos^2 \theta
\sin^2 \theta + \frac{129}{192} \sin^4 \theta) - \frac{1}{2} \la b (1-
4 \sin^2 \theta);
\nonumber \\
&=& -4 \sum_{i \neq \phi} j^{i} + \frac{3 f_{+}}{2} - 8 f_{+} \sin^2
\theta \cos^2 \theta - (\frac{1}{4} \la b + \frac{1}{8}(\partial b)^2) 
(1-4 \sin^2 \theta) \nonumber \\
&& + b^4 ( \frac{63}{64} \cos^4\theta 
- \frac{51}{32} \cos^2\theta \sin^2\theta 
+ \frac{219}{64} \sin^4 \theta) - 6Bb\sin^2 \theta. \label{r2} \\
\frac{R_{\chi \chi}}{\cos^2 \theta} &=& - \frac{1}{8} (1 - 4 \cos^2
\theta) \lgroup b (\Box^{0} - \frac{1}{6} R^{0}) b + (\partial b)^2
\rgroup + \frac{\sin \theta}{2 \cos \theta} 
\tr( (\gamma^0)^{-1} \gamma^4)_{,\theta} \nonumber \\
&& + (j^{\chi} - j^{\theta})(4 -
\frac{1}{\cos^2\theta}) - \frac{1}{2} j^{\chi}_{,\theta \theta}
- (\frac{1}{2} j^{\theta} - 2 j^{\chi} - j^{\phi})_{,\theta} 
\frac{\sin\theta}{\cos\theta} 
-  j^{\chi}_{,\theta} \frac{\cos\theta}{\sin\theta} \nonumber \\
&& + b^4 (- \frac{139}{192} \sin^4\theta + \frac{91}{96} \cos^2 \theta
\sin^2 \theta + \frac{129}{192} \cos^4 \theta) - \frac{1}{2} \la b (1-
4 \cos^2 \theta);
\nonumber \\
&=& -4 \sum_{i \neq \chi} j^{i} + \frac{3 f_{+}}{2} - 8 f_{+} \sin^2
\theta \cos^2 \theta - (\frac{1}{4} \la b + \frac{1}{8}(\partial b)^2) 
(1-4 \cos^2 \theta) \nonumber \\
&& + b^4 ( \frac{63}{64} \cos^4\theta 
- \frac{51}{32} \cos^2\theta \sin^2\theta 
+ \frac{219}{64} \sin^4 \theta) - 6Bb\cos^2 \theta. \nonumber \\
R_{r\theta} &=& -\frac{3}{2} \tr( (\gamma^0)^{-1} \gamma^4)_{,\theta}
 - 2 \sum_{i \neq \theta} j^{i}_{,\theta} + 4 (j^{\theta} - j^{\phi})
 \frac{\cos\theta}{\sin \theta} - 4 (j^{\theta} - j^{\chi})
 \frac{\sin\theta}{\cos \theta} \nonumber \\ 
&& - 2 b^4 \cos\theta \sin\theta
 (\cos^2\theta - \sin^2 \theta); \nonumber \\
&=& (b^4+2 f_{+}) \cos\theta \sin\theta (\cos^2\theta - \sin^2 \theta)
 + 2 (f_{-} - \la b) \cos\theta \sin\theta - 4 j^{r\theta}. \nonumber
\ea
Once again, we have given first the calculated curvature using the
metric ansatz and second the trace adjusted stress energy
tensor. $\Box^{0}$ is the d'Alambertian associated with covariant
derivative $\nabla^0$. Note that the d'Alambertian appears in its
conformal form. 

The only other Einstein equations which we will need are 
\ba
R_{rr} &=& - 4 \tr( (\gamma^0)^{-1} \gamma^4) - 8 \sum_{i} j^i -
\frac{3}{4} \la b + \frac{37}{36} b^4 - 3 b^4 \sin^2 \cos^2\theta 
\nonumber \\ && 
+ 4 (j^{r\theta}_{,\theta} + 2j^{r\theta} (\frac{\cos\theta}{\sin\theta} -
\frac{\sin\theta}{\cos\theta}) ); \nonumber \\
&=& 4 \tr( (\gamma^0)^{-1} \gamma^4) - 2 \alpha^{(4)} + \frac{3}{4} \la b +
 b^4 ( -\frac{83}{288} + \sin^2 \theta \cos^2 \theta) - \frac{1}{8}
 (\partial b)^2 \nonumber \\
&& - \frac{1}{2} (R^0)^{\mu\nu} R^0_{\mu\nu} + \frac{5}{36} (R^0)^2 +
\frac{1}{18} b^2 R^{0}. \label{r3} \\
(\gamma^{0})^{\mu \nu} R_{\nu \mu} &=& 8 \tr((\gamma^0)^{-1} \gamma^4)
- \frac{1}{8} \Box^{0} (b^2) + \frac{5}{72} b^2 R^{0} + \frac{1}{9}
(R^0)^2 \nonumber \\
&&+ \frac{1}{2} (R^{0})^{\mu\nu} R^0_{\mu\nu} - \frac{1}{2} 
\tr((\gamma^0)^{-1} \gamma^4)_{,\theta \theta} - 
\tr((\gamma^0)^{-1} \gamma^4)_{,\theta} (\frac{\cos \theta}{\sin
  \theta} - \frac{\sin \theta}{\cos \theta}) + 8 \sum_{i} j^{i} 
\nonumber \\
&& - \frac{3}{2} \la b + b^4 (4 \cos^2 \theta \sin^2 \theta 
- \frac{403}{288}) - 4
(j^{r\theta}_{,\theta} + 2 j^{r\theta} (\frac{\cos\theta}{\sin\theta}
- \frac{\sin\theta}{\cos\theta})); \nonumber \\
&=& 16 \tr((\gamma^0)^{-1} \gamma^4) - 8 \alpha^{(4)} -12 Bb - \la b +
\frac{353}{144}b^4 \nonumber \\
&& -2  (R^{0})^{\mu\nu} R^0_{\mu\nu} + \frac{5}{9} (R^0)^2  +
\frac{13}{72} b^2 R^{0}. \nonumber
\ea
Given the complexity of equations (\ref{eqn5}), (\ref{r2}) and
(\ref{r3}) even in the $SO(3)^2$ symmetric case we can now justify why
we considered this simpler case. Even for the ${\cal N} = 1^{\ast}$ case in
which all three mass perturbations are equal, we would have four times
as many components in (\ref{r2}) and (\ref{eqn5})!

As before, we solve these equations by expanding each unknown in
spherical harmonics
and reducing most of the differential equations to algebraic
equations. After considerable algebra we
find the following solution for metric and five-form perturbations
\ba
j^{r\theta} &=& -\frac{1}{4} \la b \cos \theta \sin\theta; \nonumber
\\
f_{-} &=& - \frac{1}{4} \la b; \nonumber \\
f_{+} &=& 12 Bb - \frac{65}{18} b^4; \label{s1} \\
\alpha^{(4)} &=& -\frac{1}{4} (R^{0})^{\mu\nu} R^0_{\mu\nu} +
\frac{1}{24} (R^0)^2 + \frac{1}{96} b^2 R^0 + \frac{1}{16} (\partial
b)^2 + \frac{1}{96} b^4; \nonumber \\
\tr((\gamma^0)^{-1} \gamma^4) &=& \frac{1}{16} (R^{0})^{\mu\nu} R^0_{\mu\nu}
- \frac{1}{72} (R^0)^2 - \frac{5}{576} b^2 R^{0} + \frac{1}{32}
(\partial b)^2 - \frac{3}{2} Bb \nonumber \\
&& + \frac{277}{576} b^4 - (12Bb + \frac{121}{36} b^4) \sin^2\theta
\cos^2 \theta. \nonumber 
\ea
Note that as anticipated the solutions depend only on $B(x^{\mu})$,
$b(x^{\mu})$ and the latter's derivatives. 
The remaining parts of the metric perturbations are such that if expand out the
angular perturbations as 
\ba
j^{\theta} &=& \lgroup \alpha \cos^4\theta + \beta \cos^2 \theta\sin^2
\theta + \gamma \sin^4 \theta \rgroup; \nonumber \\
j^{\phi} &=& \lgroup \alpha \cos^4\theta + \eta \cos^2 \theta\sin^2
\theta + \delta \sin^4 \theta \rgroup; \label{s2} \\
j^{\chi} &=& \lgroup \zeta \cos^4\theta + \xi \cos^2 \theta\sin^2
\theta + \gamma \sin^4 \theta \rgroup, \nonumber 
\ea
then 
\ba
\alpha &=& -\frac{1}{8} \la b + \frac{1}{2} Bb 
- \frac{501}{2304} b^4; \nonumber \\
\beta &=& Bb -\frac{501}{1152} b^2; \nonumber \\
\gamma &=& \frac{1}{8} \la b + \frac{1}{2} Bb - \frac{501}{2304} b^4;
\nonumber \\
\delta &=& \zeta = - \frac{195}{768} b^4; \label{s3} \\ 
\eta &=& - \frac{5}{2} Bb + \frac{527}{384} b^4 - \frac{1}{8} \la b;
\nonumber \\
\xi &=& - \frac{5}{2} Bb + \frac{527}{384} b^4 + \frac{1}{8} \la
b. \nonumber 
\ea
We now have the field perturbations to adequate order to determine 
all divergences in the action. The full solution hence depends on 
only two independent fields, $b(x^{\mu})$ and $B(x^{\mu})$, which
have a natural dual interpretation in terms of normalisable and
non-normalisable perturbations which we will discuss below in more
detail. To reinstate dimensional factors we note that $b$ has
dimensions of length and $B$ has dimensions of $(\rm{length})^3$. 
 
The remaining Einstein equations 
$R_{\mu \theta}$, $R_{\mu r}$ and $R_{\mu\nu}$ will fix the
off-diagonal elements in the metric $G_{\mu r}$ and $G_{\mu \theta}$
as well as $h^4_{\mu\nu}$. We do not need these to calculate the action
divergences and the explicit expressions are not particularly
illuminating. The only contribution to the trace of the
$h^4$ is that given in (\ref{slog}). 

Note that the Einstein equations do {\it not} fix $\gamma^4_{\mu\nu}$ 
completely. Only the trace of $\gamma^4$ and its covariant divergence
are determined. Extra data from the field theory is needed to fix the rest: the
undetermined part is specified by the expectation value of the dual
stress energy tensor \cite{HKS}. However the trace is sufficient to
determine the divergences in the action. 

Before proceeding it is useful for reference in
later sections to summarise (parts of) the solution when $b$ is
constant and $\gamma^0$ is flat
\ba
ds^2 &=& \frac{R^2}{r^2} (1 + \frac{5 b^2 r^2}{4 R^4}) 
\eta_{\mu\nu} dx^{\mu}dx^{\nu} + \frac{R^2}{r^2} dr^2 + R^2 (1+
\frac{b^2r^2}{8R^4} ) d\theta^2 \nonumber \\
&& + R^2 (1 + \frac{b^2 r^2}{8 R^4} (1 - 4 \sin^2 \theta)) d\Omega_1^2
+ R^2 (1 + \frac{b^2 r^2}{8 R^4} (1 - 4 \cos^2 \theta)) d\Omega_2^2;
\nonumber \\
\Phi &=& \ln g + \frac{b^2 r^2}{4 R^4} (\cos^2 \theta - \sin^2
\theta); \label{v1} \\
C_2 &=& \frac{1}{g} (b r + \frac{B r^3}{R^2}) 
\cos^3 \theta d\Omega_2; \nonumber
\\
B_2 &=& (b r + \frac{B r^3}{R^2}) \sin^3 \theta d\Omega_1; \nonumber
\\
C_4 &=& - (\frac{R^4}{g r^4} + \frac{b^2}{24 g r^2}) d\eta_4 + 
(\frac{4 R^4}{g} \int
\sin^2 \theta \cos^2 \theta d\theta + \frac{1}{2g} b^2 r^2 \sin^3
\theta \cos^3 \theta) d\Omega_4, \nonumber
\ea
where we have also reinstated factors of $R$ and $g$ explicitly. For
the purposes of calculating the counterterms and anomalies, we have
been considering solutions dual to field theories in general
backgrounds with position dependent masses. The latter is natural from
the supergravity perspective but is somewhat unconventional from the
field theory point of view. When we discuss the field
theory in \S \ref{six} and onwards we will restrict to the most
interesting case of flat backgrounds where the mass is constant. 

\section{The action and the conformal anomaly} \label{four}
\noindent

Having determined the series expansion of this Polchinski-Strassler
solution, we can now determine the divergent terms in the action, the
corresponding counter-terms needed to render the action finite and the
mass contributions to the conformal anomaly. 
The on-shell contribution to the volume term in the action (\ref{ac1})
is
\be
S = \frac{1}{2 \kappa^2} \int d^{10}x\sqrt{-G} \left ( \frac{1}{12}
(e^{-\Phi} H_{3}^2 + e^{\Phi} F_{3}^2) - \frac{1}{48} \bar{F}_{5}^2
\right ) - \frac{3}{4 \kappa^2} \int C_{4} \wedge H_{3} \wedge F_{3}.
\ee
Now to the order that we need
\ba
\frac{1}{12} (e^{-\Phi} H_{3}^2 + e^{\Phi} F_{3}^2) &=& 5 b^2 r^2 + 12
\la b r^4 \ln(r) \\ 
&& \hspace{2cm} 
+ r^4 (12Bb + \la b + \frac{1}{2} (\partial b)^2 - \frac{61}{16}
b^4), \nonumber 
\ea
whilst 
\ba
-\frac{\bar{F}_{5}^2}{48} &=& -8 + 3b^2 r^2 + 8 r^4 \sum_{i} j^{i} + 3
\la b r^4 (\cos^2 \theta - \sin^2 \theta) \\
&& + Bb r^4 (-12 + 96 \cos^2 \theta \sin^2 \theta) + b^4 r^4
(\frac{67}{12} -\frac{448}{9} \cos^2 \theta \sin^2 \theta) + 6 \la b
r^4 \ln(r). \nonumber 
\ea
Expanding out the metric determinant we get
\ba
\sqrt{-G} &=& \frac{\sqrt{\gamma^0}}{r^5} \sin^2 \theta \cos^2\theta
  \sin\phi\sin\chi \left ( 1 + r^2 (\frac{b^2}{48} - \frac{R^0}{12}) +
  \frac{1}{2} r^4 \sum_{i} j^i + \frac{1}{2} r^4 \tr((\gamma^0)^{-1}
  \gamma^4) \right . \nonumber \\ 
&& \left. -\frac{1}{16} (R^{0})^{\mu\nu}R^0_{\mu\nu} 
+ \frac{5}{288}(R^0)^2r^4 + \frac{1}{144} b^2 R^{0}r^4 + b^4r^4 (
\frac{1}{4} \sin^2 \theta\cos^2 \theta - \frac{427}{4608} ) \right ). 
\ea
Note that the logarithmic terms have cancelled in this expression, since
$\sum_{i} k^i + \tr((\gamma^0)^{-1} h^4) = 0$.  
Furthermore,
\ba
-\frac{3}{4 \kappa^2} \int C_{4} \wedge H_{3} \wedge F_{3} &=& 
\int d^{10}x \frac{\sqrt{\gamma^0}}{r^3} \sin^2 \theta \cos^2\theta  
\sin\phi\sin\chi \left ( -\frac{9}{2} b^2 - 18 B b r^2 - \frac{9}{2} 
\la b r^2 \right .\nonumber \\
&& \left .- 18 \la b r^2 \ln(r) 
+ \frac{3}{4} R^0 b^2 r^2 + \frac{105}{16} b^4 r^2 - 6 b^4
\sin^2\theta \cos^2 \theta. \right )
\ea
Putting all of this together, the volume term in the action is
\ba
S &=& \frac{\pi^3}{2 \kappa^2 r^5} \int dr d^4x \sqrt{-\gamma^0}
\left (  -8 + (\frac{2R^0}{3} + \frac{10b^2}{3})r^2 + r^4 (\frac{1}{4}
 (R^0)^{\mu\nu}R^0_{\mu\nu} - \frac{1}{12} (R^0)^2) \right . \nonumber \\
&& \hspace{4cm} 
\left . + \frac{3}{8} b (\Box^0b - \frac{1}{6} R^0b)r^4 + \frac{1}{24}
b^4r^4 \right ),
\ea
where we have carried out the angular integrations.  
Note that terms proportional to $B(x^{\mu})$ have cancelled as have
terms proportional to $\ln(r)/r$. The cancellation of the latter is
necessary to ensure that there is no $\ln^2 (\epsilon)$ 
divergence in the action.
It is convenient to reinstate factors of $R$ and $g$ at this stage. 
Introducing a cut-off $\epsilon$ and performing the
$r$ integration we find that the (UV) divergent terms are
\be
S = - \frac{N^2}{8 \pi^2} \int d^4x \sqrt{-\gamma^0} \left (
 \frac{2}{\epsilon^4} - (\frac{R^0}{3} + \frac{5\tilde{b}^2}{3}) 
\frac{1}{\epsilon^2} + {\cal A} \ln \epsilon \right ). \label{an}
\ee
We have switched to the natural dual
parameters, using $2 \kappa^2 = (2 \pi)^7 \alpha'^4 g^2$ and $R^4 = 4
\pi g N \alpha'^2 = g_{YM}^2 \alpha'^2$, where $g_{YM}$ is the
Yang-Mills coupling. We have also introduced what will turn out to be 
the more natural gauge theory parameter $\tilde{b}$ which is defined as 
\be
\tilde{b} = \frac{b}{\alpha' \sqrt{g_{YM}^2 N}}.
\ee 
${\cal A}$ is the conformal anomaly which is 
\be
{\cal A} = 
\frac{1}{4}
 (R^0)^{\mu\nu}R^0_{\mu\nu} - \frac{1}{12} (R^0)^2
+ \frac{3}{8} \tilde{b} (\Box^0 \tilde{b} - \frac{1}{6} R^0
\tilde{b} ) + \frac{1}{24} \tilde{b}^4.
\ee
The curvature terms are in agreement with those found in \cite{HK}
whilst the other contributions to the anomaly depend on the
field $b(x^{\mu})$. Since $B(x^{\mu})$ corresponds to a vacuum
expectation value in the dual theory it was expected that it should not
appear in the anomaly from the field theory perspective; cancellation
of the $B$ terms from the supergravity perspective was very non-trivial. 

There are additional divergent terms in the action arising from the
Gibbons/Hawking boundary term
\be
S = \frac{1}{\kappa^2} \int d^9x \sqrt{-H} {\cal K} =
- \frac{\pi^3 R^8}{\kappa^2} \int d^4x \sqrt{\gamma^0} \left (  -
\frac{4}{\epsilon^4} + (\frac{R^0}{6} - \frac{b^2}{24R^4})
\frac{1}{\epsilon^2} + .... \right ),
\ee
where ${\cal K}$ is the trace of the second fundamental form on the
induced boundary which has induced metric $H$. These divergent terms from the 
volume and surface terms should be cancelled by introducing counterterms. 
However it is not immediately apparent how we can do this here 
in a coordinate invariant way. Since the induced metric on the boundary
is nine-dimensional we cannot write a counterterm action in
terms of a four-dimensional restriction of
this such as $\gamma$ in (\ref{m0}) since this is not 
well-defined under coordinate transformations. Worse than 
this, $\gamma$ does depend on the angular coordinates explicitly. 

The best we can do in ten dimensions is the
following. Suppose we have a metric $g$ which we know
lies in the class of solutions considered here. Then we should write
$g$ in a coordinate system in which [i] it is a direct product of a
five-dimensional metric and a sphere as one approaches the boundary at
infinity and [ii] cross-terms between the two metrics fall off faster
than $\Omega^2$ where $\Omega$ is the conformal factor at the
boundary of the five-dimensional metric
in the sense of Penrose \cite{P}. This is equivalent to saying that we
should bring the metric into the coordinate system of (\ref{m0}) up
to coordinate transformations which affect only $(r,x^{\mu})$ or
the angular coordinates separately. We want the metric to be in a form
where it is manifestly asymptotic to $AdS_5 \times S^5$. 
We then identify $H$ with the metric induced on a nine-dimensional
hypersurface close to the boundary and write the counterterm action as 
\be
S_{\rm{ct}} = - \frac{R^3 }{\kappa^2} \int_{\epsilon} d^9x
  \sqrt{-H} \left (  3 + \frac{1}{4} (R(H) +
    \frac{b^2}{R^4}) \right ), \label{ct0}
\ee
where $R(H)$ is the curvature. The first two contributions 
to this action are equivalent to those 
found in \cite{BK}, \cite{KLS} and \cite{EJM}. This whole procedure
is not very satisfactory in that it is very dependent on fixing the coordinate
choice; in this sense, the counterterm action is defined much more
naturally by considering the solution in the context of
five-dimensional gauged supergravity instead. Then one has already
taken account of integrating out the angular dependence in a
natural way by implementing the dimensional reduction ansatz;
of course one still has to single out the spherical directions. 

The counterterm action (\ref{ct0}) is also very restrictive
in that it applies only to solutions which are of this specific
Polchinski-Strassler form. To work out a generally applicable counterterm
action involving matter fields would be computationally extremely difficult
since one would have to have a much more general ansatz. One can  
understand this simply from the five-dimensional perspective. Suppose
we are in a consistent subsector of the ${\cal N} = 8$ supergravity in
which only the graviton and some number $k$ of scalar fields are
switched on. Then to calculate the divergences in the action we will
have to perturbatively solve the coupled Einstein equations 
and $k$ equations for the scalars, to find (typically) $(k+1)$ terms
in both counterterm action and anomaly. As $k$
increases this process becomes more and more computationally intensive
even if $\gamma^0$ is flat. The simplification we have made here 
corresponds to switching on
only a small number of scalars in five dimensions but choosing a
particularly interesting combination from the dual perspective. 

\section{General ansatz and SL(2,R) invariance} \label{five}
\noindent

So far we have discussed the special case of the Polchinski-Strassler
solution with residual $SO(3)^2$ invariance and the RR scalar
vanishing. It turns out to be quite simple to generalise to the 
case in which the leading
order value of $C$ is non-zero and in which we switch on perturbations
of the RR scalar. 

We already have a clue as to how to construct the most general
$SO(3)^2$ invariant solution: as well
as switching on a three-form perturbation of the type (\ref{fa}) we
should also switch on a perturbation of the type (\ref{fb}). To
consistently solve the field equations we then have to allow for
non-zero perturbations of the RR scalar. 

Of course another way of constructing the general solution would be to
use the action of the $SL(2,R)$ of IIB supergravity. For example, if
one uses the S transformation $\tau \rightarrow - 1/\tau$ we go from
the solution with only $b$ non-zero to the corresponding solution with
only $a$ non-zero and interchange the spheres on which $C_2$ and $B_2$
are non-zero. However it turns out to be more illuminating to 
explicitly construct the more general solution and then interpret the
results in terms of the $SL(2,R)$.

In particular, the scalar field perturbations are
\ba
\Phi &=& \ln g + \left ( \frac{1}{4} (b^2 - a^2) r^2
+ \frac{1}{2} (\la_b b - \la_a a) r^4 \ln r \right ) 
(\cos^2 \theta - \sin^2 \theta) + .... \nonumber \\
C &=& C^0 - \left ( \frac{ab}{2g} r^2 + \frac{1}{2g} (\la_a b + \la_b a)
  r^4 \ln r \right) (\cos^2 \theta - \sin^2 \theta) + ...
\ea
where ellipses denote terms in $r^4$ which do not contribute to the
action (or field equations). $C^0$ is the background value for the RR
scalar; we have also reinstated factors of $g$ explicitly, though we
are still suppressing factors of $R$ for notational simplicity. 
The resulting solution for the three-form fluxes is the following
\ba
\bar{F}_{3} &=& - \frac{1}{g} d \left ( a  + A r^2 + F_a \sin^2 \theta
  r^2 + \la_a r^2 \ln r \right ) r \sin^3 \theta d \Omega_1 \nonumber
\\
&& + \frac{1}{g} d \left ( b  + B r^2 + F_b \cos^2 \theta
  r^2 + \la_b r^2 \ln r \right ) r \cos^3 \theta d \Omega_2 \nonumber
\\
&& + \frac{ab}{2g} r^2 (\cos^2 \theta - \sin^2 \theta) d \left (
 b \sin^3 \theta r d\Omega_1 + a \cos^3 \theta r d \Omega_2 \right
). \label{abeq} \\
H_3 &=&  d \left ( b  + B r^2 + F_b \sin^2 \theta
  r^2 + \la_b r^2 \ln r \right ) r \sin^3 \theta d \Omega_1 \nonumber \\
&& + d \left ( a  + A r^2 + F_a \cos^2 \theta
  r^2 + \la_a r^2 \ln r \right ) r \cos^3 \theta d \Omega_2. \nonumber
\ea
In these equations, 
\ba
F_a &=& - \frac{1}{2} a (a^2 + b^2); \\
F_b &=& - \frac{1}{2} b (a^2 + b^2), \nonumber 
\ea
with the following relations being satisfied
\ba
( \Box^0 - \frac{1}{6} R^0) b &=& - 2 \la_b - \frac{1}{6} b (a^2 +
b^2); \\
( \Box^0 - \frac{1}{6} R^0) a &=& - 2 \la_a - \frac{1}{6} a (a^2 +
b^2). \nonumber
\ea
There are hence now four independent fields, $\{ a,b,A,B \}$, rather
than two. Note that, since the perturbation to $C$ is non-zero, 
$\bar{F}_3$ is no
longer exact and hence one needs the non-exact term in (\ref{abeq}).

Let us consider the first two terms in the $SL(2,R)$ combination $\tau = C
+ ie^{-\Phi}$. Combining $\beta = b - ia$ then we can write
\be
\tau = (C^0 + \frac{i}{g}) - i \frac{\beta^2 r^2}{4 g} (\sin^2 \theta -
\cos^2 \theta) + ..
\ee
Thus the $a \neq 0$, $C^0 \neq 0$ solution is related to the special
solution we have already constructed by an $SL(2,R)$ transformation 
chosen such that 
\ba
\frac{i}{g} & \rightarrow &  (C^0 + \frac{i}{g}); \\
\frac{b^2}{g} & \rightarrow & \frac{\beta^2}{g}. \nonumber 
\ea
The five-form is given by
\ba \bar{F}_{\theta \Phi_4} &=& \frac{4}{g} 
\sin^2 \theta \cos^2 \theta \lbrace 1 -
\frac{3(b^2 +a^2) r^2}{8} - \frac{3 (Bb+Aa) r^4}{4} + \frac{3 f_{-} r^4}{16}
(\cos^2 \theta - \sin^2 \theta) \nonumber \\
&& + f_{+} r^4 (\frac{3}{16} -
\sin^2\theta \cos^2 \theta) + (b^2+a^2)^2 r^4 (\frac{9}{32} - \frac{1}{4}
\sin^2\theta \cos^2 \theta) \rbrace. \nonumber \\
\bar{F}_{r \Omega_{4}} &=& \frac{1}{g} 
\lbrace f_{-} - \la_b b + \la_a a + f_{+} (\cos^2 \theta
- \sin^2 \theta) -\la_a b C^0 - \la_b a C^0 
\rbrace r^3 \cos^3 \theta \sin^3 \theta. \nonumber \\
\bar{F}_{\mu \Omega_{4}} &=& \frac{1}{4g} \lbrace f_{-} + f_{+} (\cos^2 \theta
- \sin^2 \theta) \rbrace_{,\mu} r^3 \cos^3 \theta \sin^3
\theta. \label{eqn5a} \\
\bar{F}_{\theta \eta_{4}} &=& \frac{1}{g} \lbrace - f_{-} + \la_b b - \la_a a 
-\la_a b C^0 - \la_b a C^0  - f_{+} (\cos^2 \theta
- \sin^2 \theta) + 4 j^{r \theta} \rbrace \cos\theta
\sin\theta. \nonumber \\
\bar{F}_{r \eta_{4}} &=& \frac{1}{g} (\frac{4}{r^5} 
+ (\frac{(a^2+b^2)}{12 r^3} -
\frac{R^0}{3 r^3}) + \frac{\alpha^{(4)}}{r}). \nonumber
\ea
Here $f_{-}$ and $f_{+}$ are parameters which are now fixed by the
field equations in terms of the four independent variables $\{ a,b,A,B
\}$. 

The key point is that the only place in all of these field
perturbations where the background constant $C^0$ appears
is in the $\bar{F}_{r \Omega_4}$ component and its dual. 
What this means is that the solutions for the metric and five-form 
perturbations with $C^0 \neq 0$ and $a,b \neq
0$ are identical to those already presented, with the replacements in
(\ref{s1}), (\ref{s2}) and (\ref{s3}) of 
\ba
\la b & \rightarrow & \la_b b + \la_a a; \nonumber \\
B b & \rightarrow & (B b + A a); \\
b^2 & \rightarrow & (b^2 + a^2). \nonumber
\ea
The first place in the metric 
where the value of $C^0$ appears explicitly is in the perturbation
$G_{r \theta}$; this follows from the form of $\bar{F}_{r \Omega_4}$. 
However, neither of these two  perturbations will 
contribute to the divergent terms in the action. So we come to the
following conclusion. For the more general three-form perturbation given
in (\ref{abeq}), about a background in which $C^0 \neq 0$, the anomaly
in the action (\ref{an}) is given by 
\be
= \frac{1}{4}
 (R^0)^{\mu\nu}R^0_{\mu\nu} - \frac{1}{12} (R^0)^2
+ \frac{3}{8} \tilde{b} (\Box^0\tilde{b} - \frac{1}{6} R^0
\tilde{b}) + \frac{1}{24} (\tilde{b}^2 +
\tilde{a}^2)^2 +  \frac{3}{8} \tilde{a} (\Box^0\tilde{a} - \frac{1}{6}
R^0 \tilde{a}), 
\ee
whilst the counterterm action should now be taken as 
\be
S_{\rm{ct}} = - \frac{R^3 }{\kappa^2} \int_{\epsilon} d^9x
  \sqrt{-H} \left ( 3 + \frac{1}{4} (R(\gamma) +
  \frac{(a^2+b^2)}{R^4}) \right ).
\ee
Note that neither $C^0$ nor $g$ appear explicitly 
in the anomaly when written in natural dual variables: it is invariant
under $SL(2,R)$ transformations of the UV parameter $\tau$. 
As we have seen this more general case
corresponds to switching on a {\it complex} dual mass term 
$m \sim b - i a$ rather than a real mass $m \sim b$. This will become
clearer now as we consider the field theory interpretation. Note that
the anomaly depends not just on the magnitude of the mass but also
non-trivially on its phase through the derivative terms. 

\section{Dual field theory interpretation} \label{six}
\noindent

Switching on a three-form perturbation corresponds to perturbing the
dual ${\cal N} = 4$ SYM theory with mass terms. In the
language of four-dimensional ${\cal N} = 1$ supersymmetry, the ${\cal
  N} = 4$ theory consists of a vector multiplet V and three chiral
multiplets $\Phi_a$ in the adjoint representation of the gauge
group. Normalising the K\"{a}hler potential as 
\be
K = \frac{2}{g_{YM}^2} \sum_{a} {\rm{Tr}} (\bar{\Phi}_a \Phi_a), 
\ee 
then the theory has interactions summarised in the superpotential
\be 
W = \frac{2 \sqrt{2}}{g_{YM}^2} {\rm{Tr}} ( [\Phi_1, \Phi_2]
\Phi_3). \label{w1} 
\ee
Supersymmetry can be partially broken by adding mass terms to the
superpotential
\be
\Delta W = \frac{1}{g_{YM}^2} \sum_{a} m_{ab} {\rm{Tr}} \Phi_a
\Phi_b. \label{w2} 
\ee
Choosing a basis in which the mass matrix is diagonal with eigenvalues
$m_a$, then
if two masses are equal and the third is zero the theory has ${\cal N}
=2$ supersymmetry; otherwise it is the ${\cal N} = 1^{\ast}$, in the notation
of \cite{PS}. When the mass matrix vanishes, there is an $SO(6)$
R-symmetry containing an $SU(3)$ flavour symmetry under which the
$\Phi_a$ are triplets. Mass terms generically 
totally break this symmetry: $m_{ab}$ transforms in 
the ${\bar{\bf{6}}}$ of $SU(3)$. 

When the mass matrix is diagonal the F-term equations for a
supersymmetric vacuum read
\be
[\Phi_a, \Phi_b] = - \frac{m_{c}}{\sqrt{2}} \epsilon_{abc} \Phi_c.
\ee
The classical vacua of this theory were described in \cite{VW} whilst
the quantum theory was discussed in \cite{DW}. Since this was also
discussed in detail in \cite{PS} we give only a brief description here
focusing on issues relevant here.
For the purpose of describing the vacua, we can rescale the fields so
that all three masses are equal. Then the solutions to these equations
are given by N-dimensional representations of $SU(2)$, both reducible
and irreducible \cite{VW}. In the so-called Higgs vacuum the
representation is irreducible; there are also Coulomb vacua in which
the representation is the product of two or more irreducible
representations and there is an unbroken gauge group of at least
$U(1)$. 

The dual representation of these vacua discussed in \cite{PS} 
is as follows. Let us write the lowest component of $\Phi_a$ as
\be
\phi_a = (A_{a} + i \bar{A}_a),
\ee
where $A_a, \bar{A}_a$ are real. Consider a vacuum in which
\be
\phi_a = - (\frac{1}{4} \epsilon_{abc} m_b m_c)^{\frac{1}{2}} L_a,
\ee
where $L_a$ is an N-dimensional representation of $SU(2)$ satisfying
the usual commutator relation
\be
[L_a , L_b] = \epsilon_{abc} L_c. \label{com}
\ee
One now interprets the scalars $\phi_a$ as the collective
coordinates of the bulk D3-branes, normalised as $z^a = 2 \pi \alpha'
\phi_a$, where $z^a = x^{a+3} + i x^{a+6}$ written in terms of
coordinates $x^m$ on $R^6$. Suppose that the $m_a$ are real. Then,
for the Higgs phase, it was proposed that 
the branes should all lie on an ellipsoid in the $(x^4,x^5,x^6)$
directions with the radii of the ellipsoid $r_m$ determined by
\be
r_{m}^2 \equiv x^m x^m \approx \pi^2 \alpha'^2 (\frac{1}{2}
\epsilon_{abc} m_b m_c) N^2.
\ee
When the $m_a$ are complex, the ellipsoid is effectively phase 
rotated into the $(x^7,x^8,x^9)$ directions.
The idea of 
is that this system should be equivalent to a single D5-brane with
worldvolume of $M^4$ times an ellipsoid, with $N$ units of magnetic
flux through this ellipsoid \cite{My}, \cite{KT}. 
The more general Coulomb vacua are then represented by a set of D5-branes
wrapped around ellipsoids of different radii determined by the
dimensions of the irreducible representations contained in $L_a$.
When all three mass eigenvalues are equal, the branes should all lie on
spheres rather than ellipses. Thus the supergravity solution 
will have an $SO(3)$ symmetry related to the $SO(3)$ invariance of
rotations between chiral superfields. 

The supergravity story seems to be more subtle than this,
though, since it was found in \cite{PW} that the uplift of the GPPZ
flow \cite{GPPZ} gave rise to a ten-dimensional metric which had an IR
limit corresponding either to seven-branes or to five-branes on a
ring, depending on the direction in the $S^5$. One only sees
five-branes when one approaches along a direction consistent with an
IR vacuum. 

\bigskip

Now let us consider the perturbation in terms of masses for the four
Weyl fermions $\lambda_{\alpha}$
\be
\frac{1}{g_{YM}^2} m_{\alpha \beta} \lambda^{\alpha} \lambda^{\beta} +
{\rm h. c.} \label{p1}
\ee
As discussed in \cite{PS}, a diagonal mass term transforms the same
way as an imaginary-self-dual antisymmetric 3-tensor (\ref{ten}) whose
components are given in terms of the complex coordinates $z^a$ as
\ba
T &=& m_1 dz^1 \wedge d\bar{z}^2 \wedge d\bar{z}^3  + 
    m_2 d\bar{z}^1 \wedge dz^2 \wedge d\bar{z}^3  +
    m_3 d\bar{z}^1 \wedge d\bar{z}^2 \wedge d{z}^3 + \nonumber \\
&& \hspace{30mm}
    m_4 dz^1 \wedge dz^2 \wedge d{z}^3. \label{T}
\ea
The complex coordinates should be related to the coordinate system
of (\ref{met}) as
\ba
z^1 &=& x^4 + i x^7  = \frac{1}{r} 
( \cos \theta \cos \chi + i  \sin \theta \cos \phi); \nonumber \\
z^2 &=& x^5 + i x^8 = \frac{1}{r} ( \cos \theta \sin \chi \cos \psi_2 + i  
\sin \theta \sin \phi \cos \psi_1); \label{z} \\
z^3 &=& x^6 + i x^9= \frac{1}{r} ( \cos \theta \sin \chi \sin \psi_2 + i  
\sin \theta \sin \phi \sin \psi_1). \nonumber
\ea
Then we define the form 
\be
S_2 = \frac{1}{2} T_{mnp} x^{m} dx^{n} dx^p, \label{Ss1}
\ee
using the components of $T$. The leading order behaviour
of the perturbations of the bulk three-form $G_3$ defined in (\ref{Gf})
lie in the ${\bf \overline{10}}$ of $SO(6)$. The non-normalisable modes
are
\be
G_3 \sim d (\frac{R^4 r^4}{g} S_2), \label{g1}
\ee
which is of order $r$ whilst the normalisable modes are
\be
G_3 \sim d (\frac{R^6 r^6}{g} S_2), \label{g2}
\ee
which is of order $r^3$. Thus terms in $b$ in the bulk
solution correspond to mass perturbations in the field theory whilst
the terms in $B$ correspond to inducing vacuum expectation values of
the fermion bilinears $\bar{\lambda} \bar{\lambda}$. 

When all four masses are equal and non-vanishing, the $SO(4)$ symmetry
of the fermionic Lagrangian is manifest in the perturbation (\ref{p1}); the
Lagrangian is invariant under rotations of the $\lambda^{\alpha}$
which preserve $\lambda^{\alpha} \lambda^{\alpha}$. 
It is convenient to rewrite (\ref{T}) in
terms of the real coordinates for which
\be
T = 4 m (dx^4 \wedge dx^5 \wedge dx^6 - i dx^7 \wedge dx^8 \wedge
dx^9), \label{per}
\ee
which is manifestly $[SO(3)]^2$ symmetric. Furthermore, 
we can bring $T$ into the form 
\be
T = 4 m d ( \frac{1}{r^3} \cos^3 \theta d\Omega_2 - i \frac{1}{r^3}
\sin^3 \theta d \Omega_1),
\ee
which from (\ref{g1}) and (\ref{g2}) gives rise to the form for $G_3$ which 
we have been using. To fix the normalisation between $T$ and the
three-form perturbations in the bulk we can consider probes moving in
the perturbed background, which we will do in the following
section. First however we should consider a little more carefully what
the Lagrangian and the vacua of this theory are. 

When we switch on a mass for the gluino, there is no longer a
supersymmetric completion of the Lagrangian (beyond terms linear in
the mass \cite{PS}). So we need to add ``by hand'' quadratic 
mass terms for the scalars. We want to do this in such a way as to obtain 
maximal symmetry of the Lagrangian when $m_4 = m$ with the linear terms fixed by the
fermionic mass perturbation. This effectively fixes the quadratic mass
terms for the scalars: the scalar part of the potential is of the form
\be
V \propto \frac{1}{g_{YM}^2} {\rm{Tr}} \left ( \sum_{a<b} \left |
    [\phi_a,\phi_b] + \frac{m}{\sqrt{2}} \epsilon_{abc} \phi_c \right
  |^2 + \frac{2}{3 \sqrt{2}} \epsilon_{abc} {\rm{Re}}( m_4 \phi_a
  \phi_b \phi_c) + \sum_{c} \frac{\left | m_4 \right |^2}{6} \left |
    \phi_a \right |^2 \right ), \label{pot1}
\ee
where $m_4$ is the gluino mass. 
Let us again assume that the masses are real; then when the
gluino mass is equal to the other mass scale we can conveniently write
the potential in the form
\ba
\frac{1}{g_{YM}^2} {\rm{Tr}} \left ( \sum_{a<b} ([A_a,A_b]^2 +
  [A_a,\bar{A}_b]^2 + [\bar{A}_a,A_b]^2 + [\bar{A}_a,\bar{A}_b]^2) \right.
 \label{pot2} \\ 
 + \left. \frac{8m}{3 \sqrt{2}} \epsilon_{abc} A_a A_b A_c
+ \sum_{a}  \frac{2m^2}{3} (\bar{A}_a^2 + A_a^2) \right ). \nonumber
\ea
This manifestly displays an $SO(3)^2$ symmetry corresponding to
flavour rotations of the $(A_a,\bar{A}_a)$. 
The classical potential (\ref{pot2}) is extremised when the following equations
are satisfied:
\ba
2 [A_{b}, [\bar{A}_a,A_b]] + 2 [\bar{A}_b,[\bar{A}_a,\bar{A}_b]] +
\frac{4m^2}{3} \bar{A}_a = 0; \nonumber \\
2 [A_{b}, [{A}_a,A_b]] + 2 [\bar{A}_b,[{A}_a,\bar{A}_b]] +
4 \sqrt{2} m \epsilon_{abc} A_b A_c + 
\frac{4m^2}{3} {A}_a = 0, \label{z2}
\ea
where summation over repeated indices is implicit. In the trivial
vacuum in which $A_a = \bar{A}_a = 0$ the potential vanishes. We can
also find solutions for which $\bar{A}_a = 0$ and for which
\be
A_a = - \lambda L_a,
\ee
with $L_a$ defined in (\ref{com}). Then (\ref{z2}) and (\ref{pot2}) fix
\ba
\lambda_{\pm} &=& (\frac{m}{\sqrt{2}} \pm \frac{\sqrt{2}m}{3}); \\
V_{\pm} &=& \mp \frac{2m^2}{9} {\rm{Tr}} (L_a L_a). \nonumber
\ea
Now for a generic N-dimensional representation which is reducible into
$k$ irreducible representations of dimension $n_{k}$ such that $N =
\sum_{k} n_{k}$
\be
{\rm{Tr}} (L_a L_a) = \sum_{k} \frac{1}{4} n_{k} (n_{k}^2 -1). \label{dim}
\ee
This is manifestly maximised when there is a single partition, that
is, the N-dimensional representation is irreducible. So classically
the favoured configuration seems to be a Higgs phase, in which all $N$ of the
branes lie on a sphere of radius $\pi \alpha' \lambda_{+} N$ where we
have again identified the scalars $A_a$ with the locations of the
branes $x^{m}$. 

From the form of the potential (\ref{pot1}) we can show that classical
vacua in which
\be
[\phi_a,\phi_b] = - \lambda \epsilon_{abc} \phi_c,
\ee
persist in the sense that the potential is minimised at finite
$\lambda$ provided that $\left | m_4 \right | $ is less than about
$3m$. 

\section{Probe potential} \label{seven}
\noindent

In this section we will consider five-brane
probes moving in the background which we have constructed
perturbatively. If one retains only first order
corrections to the fields, this is equivalent to considering probes in
the linearised solution as in \cite{MF}. 
Let us consider probe D5-branes with worldvolume $M^4 \times S^2$ 
and D3-brane charge $n \ll N$ with $n \gg \sqrt{g N}$. 
For simplicity we will consider only D5-brane probes within the 
background of $\gamma^0$ flat, $C=0$, $a=0$ and $b$ a (real)
constant. The relevant terms in the action for the D5-branes are 
\ba
S &=& -\frac{\mu_5}{g} \int d^6\xi \left [ -{\rm{det}}(G_{M^4}) 
{\rm{det}}(g^{-\frac{1}{2}} e^{\frac{\Phi}{2}} G_{S^2} + 2 \pi \alpha'
{\cal F})  \right ]^{\frac{1}{2}} \nonumber \\
&& \hspace{20mm} 
+ \mu_5 \int ( C_{6} + 2 \pi \alpha' {\cal F}_2 \wedge C_4), \label{d5}
\ea
where
\be
2 \pi \alpha' {\cal F}_2 = 2 \pi \alpha' F_2 - B_2.
\ee
The dilaton factors follow from the fact that we are in the
Einstein as opposed to the string frame metric. 
$G_{M^4}$ is the induced metric in the $M^4$ directions of the worldvolume
and $G_{S^2}$ is the induced metrics on the two sphere.
We will choose to identify the worldvolume
coordinates $\xi^{\mu}$ with the bulk coordinates $x^{\mu}$ with
$\mu=(0,3)$.
 
The bulk potential $C_6$ is defined by the equation
\be
dC_6 - H_3 \wedge C_4 = - g^{-1} e^{\Phi} \ast{\bar{F}_3}.
\ee
This definition follows from the field equations (\ref{feq}) as well
as from the D-brane action (\ref{d5}): we require that both the action
and the field strength $\bar{F}_3$ are invariant under gauge
transformations of the type $\delta C_4 = d \chi_3$ with $\delta C_6 =
- H_3 \wedge \chi_3$ and that the field equations (\ref{feq}) are
satisfied. In our solution the potential $C_6$ is hence given by
\be
C_6 = \left [ - \frac{2b R^4}{3g r^3} - \frac{11 b^3}{24 r R^2} +
  \frac{b^3}{2 rR^2} \sin^2 \theta + ...
\right ] \sin^3 \theta d\eta_{4} \wedge d\Omega_1,
\ee
where the ellipses denote terms which are finite as $r \rightarrow
0$. Note that the second parameter in the solution, $B$, does not appear
in this potential. 

We have fixed four of the collective
coordinates of the D5-brane; let us denote the remaining six by the
(complex) fields $z^a$, with $a = (1,3)$. Then we should take the
brane configuration to be
\be
z^a = z e^a, \hspace{10mm} e^a e^a = 1,
\ee
where $e^a$ is real and parametrises the $S^2$; let us suppose that
worldvolume $S^2$ coordinates are $(\vartheta,\varphi)$. We should 
then identify $z^a$ with the bulk coordinates on the $R^6$ as in (\ref{z}).
Note that with this identification we have $\left |z \right
| = R^2 r^{-1}$.
As mentioned above, we want the D5-brane to have a D3-brane
charge of magnitude $n$. Hence we need to take a worldvolume flux of the form
\be
F_2 = - \frac{1}{2} n \sin \vartheta d\vartheta \wedge d\varphi.
\ee
The minus sign here is necessary for a D3-brane probe to remain
static since with our choice of $F_5$ the bulk solution has negative 
3-brane charge. The solutions for the bulk fields summarised in (\ref{v1}) 
along with the definition of the embedding of the brane
into the bulk give us all the ingredients necessary to expand the action
(\ref{d5}) as
\ba
S &=& - \frac{\mu_5}{g} \int d^4x \left ( \frac{2R^8}{n \alpha' r^4}
  + \frac{8 \pi b R^4}{3 r^3} \sin^3 \theta 
+ \frac{2 \pi^2 n b^2 \alpha'}{3 r^2} \right .\label{5br} \\
&& \hspace{20mm} \left. + \frac{\pi b^3}{ r} \sin^3 \theta (\frac{5}{3} - 2
\sin^2 \theta) + \frac{\pi^2 b^4 n \alpha'}{12 R^4} \ln(r/R) + ..... 
\right ), \nonumber 
\ea
where we have integrated out the worldvolume 
angular dependence and the ellipses
denote terms which are finite or tend to zero as $r \rightarrow 0$. We
have used the conditions that $n \ll N$ and $n \alpha' \gg R^2$ to
retain only the leading order coefficient at each power of $r$. At
leading order there is a large cancellation between the Born-Infeld
and Chern-Simons terms, since a D3-brane probe feels no force from
D3-branes. The perturbations of $B_2$ and $\phi$ contribute at a 
subleading level to the action and only affect terms of order $1/r$ or
smaller. Notice that the second parameter in the bulk solution, 
$B$, does not appear in this action at all. 

We would now like to match this action to the potentials discussed in
the previous section. To do this, we should first introduce a new 
radial coordinate $\rho$ such that 
\be
\rho = \frac{1}{r} \left [ 1 +
  \frac{\tilde{b}^2 r^2}{24} (5 - 6 \sin^2 \theta) 
+ \frac{\pi^2 \tilde{b}^4 r^4 \ln(r)}{96 g_{YM}^2 N} 
+ ...\right ], \label{c1}
\ee 
where we have switched to the natural gauge theory parameters,
$\tilde{b}$, introduced previously, and $g_{YM}$; 
ellipses denote terms which are finite or zero as $r \rightarrow 0$. 
This rescaling absorbs the last two divergent terms in
(\ref{5br}). Both (\ref{c1}) and (\ref{5br}) are applicable only when
$r \tilde{b} < 1$ so that our perturbative construction of the bulk
solution is valid.

Now we should relate the gauge theory scalar $\phi$ to
the D-brane collective coordinates as
\be 
\phi = \frac{\sqrt{g_{YM}^2N}}{\sqrt{2} \pi} \rho e^{i\theta}. \label{phz}
\ee
With this identification, the probe action becomes
\be
S =  - \frac{16}{n g_{YM}^2} \int d^4x \left [ \left|\phi \right |^4 - \frac{
    \tilde{b} n}{6 \sqrt{2}} (3 \left | \phi \right |^2
  {\rm{Im}}(\phi) + {\rm{Im}}(\phi)^3) + \frac{\tilde{b}^2
    n^2}{24} \left | \phi \right| ^2 \right ], \label{na} 
\ee
where we have used $\mu_5^{-1} = (2 \pi)^5 \alpha'^3$. 
To make a comparison with the field theory, we should return to the
potential (\ref{pot1}). Let us make the matrix scalar fields to be of
the form 
\be
\phi_a = - \frac{2i}{n} \phi L_a,
\ee
where $\phi$ is a scalar and $L_a$ is the $n$-dimensional
irreducible representation of $SU(2)$. Substituting into (\ref{pot1})
and making use of (\ref{dim}) the scalar potential is 
\be
\frac{4}{n g_{YM}^2} \left ( \left | \phi \right |^4 + \frac{n}{3
    \sqrt{2}} (3 \left | \phi \right |^2 {\rm{Im}}(m \phi) +
  {\rm{Im}}(m \phi^3) ) + \frac{n^2}{6} \left | m \right |^2 \left |
    \phi \right |^2 \right ),
\ee
which agrees with the form (\ref{na}) when $m$ is real. It was
mentioned in \cite{PS} that there is an ambiguity in the quadratic
term of this potential: we can add another term corresponding to
traceless combination of masses for the scalar bilinears. We chose
this to vanish in (\ref{pot1}) but even if we had not we could have
absorbed it in the matching between gauge theory scalar and collective
coordinate. Comparison with (\ref{na}) then gives the identification
\be
m \equiv - \frac{\tilde{b}}{2},
\ee
which means that the anomaly from the mass perturbation is
\be
S = - \frac{N^2}{8 \pi^2}\int  d^4x \sqrt{-\gamma^0} \left ( 
\frac{3}{2} m (\Box^{0} m - \frac{1}{6} R^{0} m) + \frac{2}{3} m^4
\right ). \label{an1}
\ee
We will interpret this in the next section. 
Now returning to the probe action (\ref{na}) we see that in addition
to the extremum at $\phi = 0$ there will also be an extremum when 
\be
\phi = \phi_e = \frac{i n \tilde{b}}{2 \sqrt{24}} ( \sqrt{3} + 1),
\label{ph} 
\ee
which is at an absolute minimum of the potential. Thus it seems that is
energetically favourable for the $D5$-brane to sit at finite radius in
the bulk. To check this we need to match the gauge theory scalar back
to the collective D-brane coordinate and consider carefully the
range of validity of our bulk solution. From (\ref{c1}), (\ref{ph})
and (\ref{phz}) the coordinate position of the minima is at
\be
\tilde{b} r_{e} \approx \frac{\sqrt{g N}}{n} \ll 1,
\ee
at which point our solution is still valid. More generally we could
consider $(c,d)$ probes and find their minima: for example, 
we expect that an NS5-brane should
sit at $\theta = 0$ at a radius approximately $g$ smaller than the
D5-brane minima. 

\section{Conformal anomalies in the field theory} \label{eight}
\noindent

The addition of mass terms breaks conformal
invariance and so the stress energy tensor is not traceless even
classically. For example if we consider free massive scalar
fields in a flat background:
\be
S = \int d^4 x \sqrt{\eta}
( (\partial \phi)^2 + m^2 \phi^2), \label{op}
\ee
then the action is manifestly not invariant under conformal
transformations of the metric $\eta_{\mu\nu} \rightarrow 
\Omega^2 \eta_{\mu\nu}$. 
On dimensional grounds the anomalous scaling of the action under
conformal transformations must be
\be
S \sim m^4 \ln \epsilon \int d^4 x,
\ee
where $\epsilon$ has dimensions of length. Now let us consider 
the coefficient in the mass perturbed YM theory.
As usual calculation of the
coefficient in the regime of large t'Hooft coupling where the
supergravity calculations are valid is inaccessible. However, it is
straightforward to calculate the coefficients in the weak coupling
(free field) regime.
Naively there is no reason why the weak and strong coupling coefficients
should agree. However, like the curvature part of the conformal
anomaly \cite{HK}, the mass anomaly term depends only on $N$ and the
UV mass parameter $m$ and does not involve the coupling
explicitly. This is indicative that the coefficient may be reproduced
by a weak coupling calculation and is not renormalised. 

So let us now determine the coefficient for massive multiplets in
the adjoint representation of $SU(N)$ in the zero coupling limit in a
flat background. To do this, we can use zeta function regularisation 
to calculate the action for free massive scalars and Weyl spinors. 
Both scalar and spinor results are well-known and can be determined
very easily. For example, for the scalars, 
the eigenvalues of the differential operator
associated with (\ref{op}) (in Euclidean spacetime) are
\be
\lambda_k = k^2 - m^2.
\ee
Then if we define the generalised zeta function as $\zeta(s) =
\sum_{k} (\lambda_k)^{-s}$ the anomalous scaling of the associated 
Euclidean action is given by
\be
S_{\rm{anomalous}} = - \ln(\epsilon) \zeta(0),
\ee
where $\epsilon$ is the scale. Here we have
\be
\zeta(s) = \frac{V}{(2\pi)^4} \int d^4k (k^2 - m^2)^{-s} = 
\frac{Vm^4}{(2\pi)^4} \int d^4u (u^2 - 1^2)^{-s},
\ee
where $V$ is the volume of the spacetime, given by the regularisation
of $\int d^4x \sqrt{\eta}$, and in the second inequality we 
explicitly demonstrate
the dimensions. Carrying out the integral above then gives
the anomalous part of the scalar action to be
\be
S_{\rm{scalar}} = - \frac{m^4}{16 \pi^2} \ln\epsilon \int d^4 x , \label{sc}
\ee
for a free (complex) scalar. This is in agreement with the result
obtained in \cite{CW} and \cite{IIM}. Both the scalar result and the
anomalous part of the action for Weyl spinors
\be
S_{\rm{spinor}} = \frac{m^4}{16 \pi^2} \ln \epsilon \int d^4x, \label{sp}
\ee
can be found in the review article \cite{BCVZ}. 

For the ${\cal N} =0$ theory with all four fermionic masses equal to
$m$, the three complex scalars acquire masses of $\sqrt{4m^2/3}$: this
follows from the potential (\ref{pot1}).
Since the fields transform in the adjoint representation of $SU(N)$, in the
free field limit the anomalous part of the action is given by
\ba
S_{\rm{anomalous}} &=& (N^2 -1) (S_{\rm{scalar}} + S_{\rm{spinor}});
\nonumber \\
&=& (N^2 -1) ( - \frac{m^4}{3\pi^2} + \frac{m^4}{4 \pi^2}) 
\ln \epsilon \int d^4x =
- \frac{m (N^2-1)}{12 \pi^2} \ln \epsilon \int d^4x,
\ea
which is in agreement with (\ref{an1}) in the large $N$ limit! 
Given this agreement 
one would expect that a calculation along the lines of \cite{CJ},
\cite{CE}, \cite{DDI} and \cite{CDJ} would also reproduce the mixed
parts of the anomaly in (\ref{an1}) for a position dependent mass term.

In the ${\cal N} = 4$ theory, coefficient of the gravitational 
conformal anomaly 
is not renormalised; agreement between strong and weak coupling 
results is an indication of the existence of
the non-renormalisation theorems discussed in
\cite{AFGJ}, \cite{GK}, \cite{HSW}, \cite{PKS}. Here however the
theory is no longer supersymmetric or conformal even classically and
so we would not expect such theorems to go through. Since conformal
anomalies are intimately connected with UV renormalisation and our
theory flows to the ${\cal N} = 4$ theory in the UV, the
non-renormalisation of the mass anomaly must be inherited from
the conformal theory. We leave as an open question the issue of making
this statement more precise. 

\section{General comments on Polchinski-Strassler solutions} \label{nine}
\noindent

In this section we will make a number of comments about the full form
for this Polchinski-Strassler solution and about the general solutions
with unequal fermion masses. 

The range of validity of the solution which we have constructed is
$\tilde{b} r \ll 1$; as we go further into the interior the
perturbative series breaks down. To probe the IR physics in the
field theory, we would have to
construct the full bulk solution or at least match our asymptotic solution
to some interior near-brane solution. Obviously to find the IR
structure of our solution would be very interesting.
Since the bulk solution has $SO(3)^2$ symmetry, all fields depend only
on two coordinates and it may be tractable to construct the full
solution in ten dimensions, using the asymptotic solution for
guidance. For example, the two-forms must be of the form
\be
B_2 = B(r,\theta) d\Omega_1; \hspace{10mm} C_2 = C(r,\theta)
d\Omega_2,
\ee
and the asymptotic solution will guide us as to appropriate ans\"atze
for the other fields. However, since no supersymmetry is preserved 
we would have to solve the full set of second-order equations which
would be challenging.

Another way to obtain insight about the full solution would be to try
a matching to a near shell metric along the lines of that in
\cite{PS}. Consider the near shell
solution obtained by simply replacing the harmonic 
function in (\ref{bas}) with that for 3-branes distributed on a shell of
radius $w$ 
\be
H = \frac{R^4}{(x^2 + 2 x w \cos \tilde{\theta} + w^2)
(x^2 - 2 x w \cos \tilde{\theta} + w^2)},
\ee
where $x$ is the AdS radius and $\tilde{\theta}$ is an angle such that
the branes are located on the sphere at $\tilde{\theta} =0$. This
metric cannot be matched even to first order to the asymptotic solution we have
constructed here, by any matching between $(x,\tilde{\theta})$ and
$(r,\theta)$, and is hence not the appropriate near-shell solution. 

This is understandable from the supergravity point of view 
in that as we flow into the interior the 3-form
perturbations become significant and so we do not expect the leading
order metric to be the same as in the $G_3 = 0$
case even if the correct interpretation is in terms of 3-branes 
distributed on a sphere.  
In fact given the work of \cite{PW} we should not expect
that the IR solution looks like five-branes wrapped on a sphere. It is
likely that the IR physics is far more subtle: the supergravity
approximation probably fails at some scale short of the infra-red. It
is also quite possible that the flow runs into a ``brick wall'' as in
\cite{PW}. 

Since there is a high degree of symmetry in our solution, it might 
be profitable to try and construct the full solution in five
dimensions and then uplift it. It is somewhat ironic, given the
motivation of this paper, that in the example specifically considered
here it is probably easier to work with the five-dimensional fields.
Unfortunately since there is no supersymmetry
we will have able to use the Einstein equations rather than the by now familiar 
first order equations for flows involving the superpotential \cite{DGPW}. 
However, this is probably still the most tractable method of obtaining
the full solution, as we will only need to switch on
a small subset of the scalars and the graviton in five dimensions. 

This follows from restricting to the $[SO(3)]^2$ invariant sector of the
scalar manifold of gauged ${\cal N} = 8$ supergravity in five
dimensions: the resulting manifold will be only four-dimensional. The
four scalars are dual to the gauge coupling, the theta angle and a
single complex fermion bilinear
\be
{\cal{O}} = \sum_{a=1}^{4} {\rm{Tr}} (\lambda_a \lambda_a).
\ee
Following \cite{GPPZ} and \cite{PW2} it is probably consistent to keep
only the latter. Then if $\sigma$ is the corresponding supergravity
scalar the scalar potential will be of the form
\be
V = - \frac{2}{R^2} \left (4 + e^{-\sigma} + 
e^{\sigma} \right), \label{nv}
\ee
where $\sigma$ is canonically normalised. This form for the potential
follows from restricting the $SO(3)$ invariant potentials of
\cite{PW2}, \cite{KPW} further. 
The only critical point of this potential is at $\sigma = 0$,
corresponding to the UV conformal fixed point: we 
expect this since there is no critical point preserving $SO(3)^2$
invariance \cite{KPW}. Thus the ${\cal N} =0$ flow is a flow to Hades
in the sense of \cite{DGPW} and both the five and the ten dimensional
solutions are going to be singular in the IR. With a one-dimensional
potential solving for five-dimensional flows may be tractable.
Even more subtle, however, will be uplifting the solution
to ten dimensions and correctly matching higher and lower dimensional
fields. We hope to report on this elsewhere.

\bigskip

Our second group of comments relate to switching on general fermion masses,
discussed in the context of five-dimensional flows in \cite{Wa2}. 
An interesting insight into the structure of the unequal mass
solutions comes from considering the linearised dilaton perturbations
induced by switching on the three-forms. These are particularly easy
to construct since they are independent of the linearised metric 
perturbations. Let us assume that the masses are real so as to prevent
source terms for the RR scalar field. Then the dilaton equation of
motion is just
\be
D^{M} \partial_{M} \Phi = \frac{g^2}{24} \left [ G_{mnp} G^{mnp} +
  {\rm c.c} \right ].
\ee
Using the explicit forms for $G$ from (\ref{T}), (\ref{Ss1}) and
(\ref{g1}), we find the following simple result
\be
D^{M} \partial_{M} \Phi = 8 r^2 (m_1 m_4 + m_2 m_3) (\sin^2 \theta -
\cos^2 \theta),
\ee
which leads to the linearised solution
\be
\Phi = \ln g + \frac{1}{2} (m_1 m_4 + m_2 m_3) r^2 (\cos^2 \theta -
\sin^2 \theta).
\ee
There are various observations to make about this. Firstly, the masses
do not appear symmetrically as one might have naively expected. The
masses will also not appear symmetrically in either the counterterms
or the anomaly. The anomaly in fact vanishes except when $m_4 \neq
0$. We can prove this from both the field theory and the
supergravity perspective. From the field theory point of view,
if the anomaly is given by that in the free field limit, then it 
must vanish unless $m_4 \neq
0$ since the contribution from each massive chiral superfield is zero
from (\ref{sc}) and (\ref{sp}). 

However, the supergravity
result can also be derived from the work of \cite{MMT2} and
\cite{GPPZ}. A result given in \cite{MMT2} is that there is no anomaly 
if the scalar potential for a canonically normalised scalar $\varphi$ 
satisfies the following conditions at the critical point $\varphi = 0$
\be
V = -\frac{12}{R^2}; \hspace{5mm} 
\frac{\partial^2 V}{\partial^2 \varphi} = -\frac{8}{R^2}
\ee
The GPPZ potential which corresponds to equal masses for three 
fermions (and zero gluino condensate) indeed
satisfies this condition and so the anomaly vanishes. (Note that 
the potential (\ref{nv}) does not satisfy this condition permitting a
non-zero anomaly.) Hence we have further evidence for 
non-renormalisation of the mass anomaly. 

Secondly, although in general the symmetry group is totally broken
down, when the masses are real the first correction to the dilaton
depends only on the angle $\theta$. When the masses are complex, the
dilaton depends on more angular coordinates. Also second order
corrections which become significant as one flows further into the
interior probably depend on more of the $S^5$ coordinates, though we
would have to go beyond the linearised level to show this. 

Thirdly, if we switch off the mass perturbations but retain
expectation values for the fermion bilinears the leading order
corrections to the dilaton vanish. This is apparent from the form of
(\ref{g2}) and in particular using (\ref{v1}). From the supergravity
point of view this is not obvious a priori but in the field theory we do not
expect the coupling to run unless masses are switched on \cite{PW2}, 
\cite{PZ}. The vacuum expectation value is not expected to affect the
UV physics, as we have found here, but to crucially determine 
the IR behaviour.

Finally, we come to the most significant point: the
coordinate dependence of the dilaton (or more generally the complex
coupling $\tau$). The flow of the dilaton should be interpreted in 
terms of the scale dependence of the field theory gauge coupling. 
However, this interpretation is not easy to make precise. Non-trivial
operator mixings may cause the dilaton to become some other coupling
of the effective action as one flows towards the IR. Also 
the dilaton does not depend just on the UV mass scales and 
the radius, to be interpreted as the energy
scale in the dual theory: it also depends upon the bulk angular
coordinates. This was commented on in \cite{MF}, \cite{PW} and 
was also discussed in the context of ${\cal N} =2$ flows in \cite{PW2}. 
The flow knows about which directions are getting massive; however, to
make sense of this we
would need to understand how this direction is encapsulated in the
effective action and how it is seen from the brane perspective.

\acknowledgements
This work is funded by St John's College, Cambridge and by the PPARC
SPG programme ``String theory and realistic field theory'',
PPA/G/S/1998/00613. The author would like to
thank Princeton University for hospitality during the completion of
this work.


\begin{thebibliography}{}

\bibitem{Ma} J. Maldacena, Adv. Theor. Math. Phys. {\bf 2} (1998)
  231, hep-th/9711200.

\bibitem{GKP} S. Gubser, I. Klebanov and A. Polyakov, Phys. Lett. {\bf
    B428} (1998) 105, hep-th/9802109.

\bibitem{W1} E. Witten, Adv. Theor. Math. Phys. {\bf 2} (1998) 253,
  hep-th/9802150. 

\bibitem{A} O. Aharony, S. Gubser, J. Maldacena, H. Ooguri and Y. Oz,
  Phys. Rep. {\bf 323} (2000) 183, hep-th/9905111. 

\bibitem{BK} V. Balasubramanian and P. Kraus, Commun. Math. Phys. {\bf
    208} (1999) 413, hep-th/9902121.

\bibitem{KLS} P. Kraus, F. Larsen and R. Siebelink, Nucl. Phys. {\bf
    B563} (1999) 259, hep-th/9906127.

\bibitem{GH} G. Gibbons and S. Hawking, Phys. Rev. {\bf D15} (1977)
  2752. 

\bibitem{LT} H. Liu and A. Tseytlin, Nucl. Phys. {\bf B533} (1998) 88,
  hep-th/9804083.

\bibitem{HK} M. Henningson and K. Skenderis, JHEP {\bf 9807} (1998) 023,
  hep-th/9806087; hep-th/9812032.

\bibitem{EJM} R. Emparan, C. Johnson and R. Myers, Phys. Rev. {\bf
    D60} (1999) 104001, hep-th/9903238.

\bibitem{MMT1} M. Taylor-Robinson, hep-th/0001177.

\bibitem{MMT2} M. Taylor-Robinson, hep-th/0002125.

\bibitem{W2} E. Witten, Adv. Theor. Math. Phys. {\bf 2} (1998) 505,
  hep-th/9803131. 

\bibitem{R1} J. Russo, Nucl. Phys. {\bf B543} (1999) 183,
  hep-th/9808117.

\bibitem{C1} C.~Csaki, Y.~Oz, J.~Russo and J.~Terning, Phys. Rev. 
{\bf D59} (1999) 065012, hep-th/9810186.

\bibitem{R2} J.~G.~Russo and K.~Sfetsos,
Adv.\ Theor.\ Math.\ Phys.\  {\bf 3} (1999) 131,
hep-th/9901056.

\bibitem{Cv1}
M.~Cvetic and S.~S.~Gubser, JHEP {\bf 9904} (1999) 024, hep-th/9902195.

\bibitem{Cv2} M.~Cvetic and S.~S.~Gubser, JHEP {\bf 9907} (1999) 010,
hep-th/9903132.

\bibitem{K1}
I.~R.~Klebanov and A.~A.~Tseytlin, Nucl. Phys.  {\bf B546} (1999) 155,
hep-th/9811035.

\bibitem{M1}
J.~A.~Minahan, JHEP {\bf 9901} (1999) 020, hep-th/9811156.

\bibitem{K2}
I.~R.~Klebanov and A.~A.~Tseytlin, Nucl.\ Phys.\  {\bf B547} (1999) 143,
hep-th/9812089.

\bibitem{M2}
J.~A.~Minahan, JHEP {\bf 9904} (1999) 007,
hep-th/9902074.

\bibitem{KS1} A.~Kehagias and K.~Sfetsos, Phys.\ Lett.\  
{\bf B454} (1999) 270, hep-th/9902125.

\bibitem{G1} S.~S.~Gubser, hep-th/9902155.

\bibitem{KS2} A.~Kehagias and K.~Sfetsos,
Phys.\ Lett.\  {\bf B456} (1999) 22, hep-th/9903109.

\bibitem{GPPZ} L. Girardello, M. Petrini, M. Porrati and A. Zaffaroni,
  Nucl. Phys. {\bf B 569} (2000) 451, hep-th/9909047.

\bibitem{PS} J. Polchinski and M. Strassler, hep-th/0003136.

\bibitem{Ar1} I. Klebanov and E. Witten, Nucl. Phys. {\bf B 536} (1998)
  199, hep-th/9807080.

\bibitem{Ar2} I. Klebanov and N. Nekrasov, Nucl. Phys. {\bf
  574} (2000) 263, hep-th/9911096.

\bibitem{Ar3} I. Klebanov and A. Tseytlin, Nucl. Phys. {\bf B578}
  (2000) 123, hep-th/0002159.

\bibitem{Ar4} I. Klebanov and M. Strassler, JHEP {\bf 0008}
  (2000) 052, hep-th/0007191.

\bibitem{Ar5} L. Pando Zayas and A. Tseytlin, JHEP {\bf 0011} (2000)
  028, hep-th/0010088.

\bibitem{Ar6} L. Pando Zayas and A. Tseytlin, hep-th/0101043.

\bibitem{Ar7} A. Buchel, C. Herzog, I. Klebanov. L. Pando Zayas and
  A. Tseytlin, hep-th/0102105. 

\bibitem{PW} K. Pilch and N. Warner, hep-th/0006066. 

\bibitem{Wa1} N. Warner, hep-th/0011207.

\bibitem{Wa2} A. Khavaev and N. Warner, Phys. Lett. {\bf B495} (2000)
  215, hep-th/0009159.

\bibitem{My} R. Myers, JHEP {\bf 9912} (1999) 022, hep-th/9910053.

\bibitem{NV} H. Nastase and D. Vaman, Nucl. Phys. {\bf B583} (2000)
  211, hep-th/0002028.

\bibitem{PW2} K. Pilch and N. Warner, Nucl. Phys. {\bf B594} (2001)
  209, hep-th/0004063.

\bibitem{PW3} K. Pilch and N. Warner, Phys, Lett. {\bf 487B} (2000)
  22; hep-th/0002192.

\bibitem{CLPST} M. Cvetic, H. Lu, C. Pope, A. Sadrzadeh and T. Tran, 
 Phys. Rev. {\bf D62} (2000) 064028, hep-th/0003103. 

\bibitem{GrP} M. Grana and J. Polchinski, Phys. Rev. {\bf D63} (2001)
  026001, hep-th/0009211.

\bibitem{MF} D. Freedman and J. Minahian, JHEP {\bf 0101} (2001) 036,
  hep-th/0007250. 

\bibitem{BBO} E. Bergshoeff, H. Boonstra and T. Ortin, Phys. Rev. {\bf
    D53} (1996) 7206, hep-th/9508091.

\bibitem{Sch} J. Schwarz, Nucl. Phys. {\bf B226} (1983) 269.

\bibitem{PH} P. Howe and P. West, Nucl. Phys. {\bf B238} (1984) 181. 

\bibitem{GL} C. Graham and J. Lee, Adv. Math. {\bf 87} (1991) 186.

\bibitem{FG} C. Fefferman and C. Graham, ``Conformal Invariants'', in 
 {\it Elie Cartan et les Math\'{e}matiques d'aujourd'hui}
 (Ast\'{e}risque, 1985) 95. 

\bibitem{FGPZ} D. Freedman, S. Gubser, K. Pilch and N. Warner, JHEP
  {\bf 0007} (2000) 38; hep-th/9906194.

\bibitem{DGPW} D. Freedman, S. Gubser, K. Pilch and N. Warner,
  hep-th/9904017; JHEP {\bf 0007} (2000) 023, hep-th/9906194.

\bibitem{PZ} M. Petrini and A. Zaffaroni, hep-th/0002172.

\bibitem{P} R. Penrose and W. Rindler, {\it Spinors and Spacetime},
  volume 2, chapter 9 (Cambridge University Press, Cambridge, 1986). 

\bibitem{HT} M. Henneaux and C. Teitelboim, Phys. Lett. {\bf B142}
  (1984) 355.

\bibitem{H} S. Hawking, Phys. Lett. {\bf B126} (1983) 175.

\bibitem{HKS} S. de Haro, K. Skenderis and S. Solodukhin, hep-th/0002230.

\bibitem{VW} C. Vafa and E. Witten, Nucl. Phys. {\bf B431} (1994) 3;
  hep-th/9408074. 

\bibitem{DW} R. Donagi and E. Witten, Nucl. Phys. {\bf B460} (1996)
  299; hep-th/9510101.

\bibitem{KT} D. Kabat and W. Taylor, Phys. Lett. {\bf B426}
  (1998) 297, hep-th/9712185. 

\bibitem{CW} S. Coleman and E. Weinberg, Phys. Rev. {\bf D7} (1973) 1888.

\bibitem{IIM} J. Iliopoulos, C. Itzykson and A. Martin,
  Rev. Mod. Phys. {\bf 47} (1975) 165.

\bibitem{BCVZ} A. Bytsenko, G. Cognola, L. Vanzo and S. Zerbini,
  Phys. Rept. {\bf 266} (1996) 1, hep-th/9505061. 

\bibitem{CJ} S. Coleman and R. Jackiw, Ann. Phys. {\bf 67} (1971) 552.

\bibitem{CE} M. Chanowitz and J. Ellis, Phys. Rev. {\bf D7} (1973) 2490.

\bibitem{DDI} S. Deser, M. Duff and C. Isham, Nucl. Phys. {\bf B111}
  (1976) 45.

\bibitem{CDJ} J. Collins, M. Duncan and S. Joglekar, Phys. Rev. {\bf
    D16} (1977) 438.

\bibitem{AFGJ} D. Anselmi, D. Freedman, M. Grisaru and A. Johansen,
  Nucl. Phys. {\bf B 526} (1998) 543, hep-th/9708042; D. Anselmi,
  J. Erlich, D. Freedman and A. Johansen, Phys. Rev. {\bf D57} (1998)
  7570, hep-th/9711035.

\bibitem{GK} S. Gubser and I. Klebanov, Phys. Lett. {\bf B413} (1997)
  41, hep-th/9708005.

\bibitem{HSW} P. Howe, E. Sokatchev and P. West, Phys. Lett. {\bf
    B444} (1998) 341, hep-th/9808162.

\bibitem{PKS} A. Petkou and K. Skenderis, Nucl. Phys. {\bf B561} (1999)
  100, hep-th/9906030. 

\bibitem{KPW} A. Khavaev, K. Pilch and N. Warner, Phys. Lett. {\bf
    B487} (2000) 14, hep-th/9812035. 

\end{thebibliography}
\end{document}